\documentclass[onecolumn,a4paper,12pt,oneside]{article} 
\usepackage[top=2.5cm,left=2.5cm,right=2.5cm,bottom=3cm]{geometry}    
\usepackage[nostamp]{draftwatermark} 
\SetWatermarkText{\bf DRAFT}
\SetWatermarkAngle{45}
\SetWatermarkScale{4}
\SetWatermarkLightness{0.85}     
\usepackage[parfill]{parskip}    
\usepackage{graphicx}
\usepackage{amssymb}
\usepackage{amsmath}
\usepackage{amsfonts}
\usepackage{amsthm}
\usepackage{epstopdf}
\usepackage[usenames,dvipsnames,svgnames,table]{xcolor}
\usepackage{tikz}
\usepackage{array}
\usepackage[T1]{fontenc}
\usepackage[utf8]{inputenc}
\usepackage{booktabs} 
\usepackage{array} 
\usepackage{paralist} 
\usepackage{listings} 
\usepackage{verbatim} 
\usepackage{subfig} 
\usepackage[colorlinks=true]{hyperref}
\hypersetup{colorlinks=true, linktocpage=true, linkcolor=Blue, citecolor=Green,
hypertexnames=true, urlcolor=MidnightBlue,pdftex}
\usepackage[english]{babel}
\usepackage{wrapfig}
\usepackage{multirow}
\usepackage{quoting}
\usepackage{rotating}
\quotingsetup{font=normalsize}
\theoremstyle{plain}

\usepackage{bm}
\usepackage{abstract}
\usepackage{float}
\usepackage{textcomp} 	
\usepackage{tabularx} 
\usepackage{caption}
\usepackage{authblk} 
\usepackage{eso-pic} 
\usepackage{enumitem} 
\usepackage{lineno} 
\providecommand{\keywords}[1]{\textbf{{Key words: }} #1} 

\usepackage{moreverb,url}

\newcommand{\be}{\begin{equation}}
\newcommand{\ee}{\end{equation}}
\newcommand{\bs}{\begin{split}}
\newcommand{\es}{\end{split}}

\newcommand{\beps}{\boldsymbol{\varepsilon}}

\newcommand{\bk}{\boldsymbol{\kappa}}

\renewcommand{\Phi}{\varPhi}
\newcommand{\gr}{\mathbf}

\renewcommand{\Theta}{\varTheta}
\renewcommand{\Psi}{\varPsi}
\renewcommand{\Sigma}{\varSigma}
\newcommand{\A}{\mathbb{A}}
\newcommand{\Ac}{\mathcal{A}}
\newcommand{\B}{\mathbb{B}}
\newcommand{\Bc}{\mathcal{B}}

\newcommand{\D}{\mathbb{D}}
\newcommand{\Dc}{\mathcal{D}}
\newcommand{\Q}{\mathbb{Q}}
\renewcommand{\H}{\mathbb{H}}
\renewcommand{\Delta}{\varDelta}
\renewcommand{\phi}{\varphi}
\renewcommand{\psi}{\varPsi}

\newcommand{\N}{\mathbf{N}}
\newcommand{\M}{\mathbf{M}}

\begin{document}

\title{{\bf About elastic coupled anisotropic laminates}\bigskip\\}

\author{Paolo Vannucci\\ 
\begin{small}LMV - Laboratoire de Mathématiques de Versailles, UMR8100.\\Université de Versailles et Saint Quentin - \href{mailto:paolo.vannucci@uvsq.fr}{paolo.vannucci@uvsq.fr}\end{small}\medskip\\--------------------\medskip\\
\small{Preprint of : On the mechanical and mathematical properties of the stiffness and compliance coupling tensors of composite anisotropic laminates. Journal of Composite Materials. 2023;57(26):4197-4214. \href{https://journals.sagepub.com/doi/10.1177/00219983231206600}{doi:10.1177/00219983231206600 }}}

\date{}

\maketitle

\section*{Abstract}
This paper contains a set of theoretical results concerning the coupling tensor $\B$ of an anisotropic laminate and of its compliance corresponding $\Bc$. The theoretical analysis and the mechanical results are obtained through an extensive use of the so-called polar formalism, introduced as early as 1979 by Prof. G. Verchery.

\keywords{anisotropy, laminates, bending-extension coupling, polar formalism}

\section{Introduction}
We consider in this paper some peculiarities of coupled anisotropic laminates. In particular, we investigate the coupling stiffness, $\B$, and compliance, $\Bc$, tensors, their algebraic structure and properties, their relations to the stiffness extension and bending tensors, respectively $\A$ and $\D$, and their influences on the compliance extension and bending tensors, $\Ac$ and $\Dc$ in the order.  The investigation is carried on using the so-called polar formalism, a mathematical technique introduced as early as 1979 by Professor G. Verchery,  \cite{verchery79,meccanica05,vannucci_libro}. The true advantage of the polar method is that a planar tensor of any order is represented by invariants and angles, which is very useful when dealing with anisotropic problems, where by definition everything depends on the direction. Moreover, in the polar formalism all the material symmetries are represented by special values of some tensor invariants. 

In the paper, after recalling the essentials of the Classical Laminated Plates Theory (CLPT) and a brief introduction to the polar formalism for laminates, we first consider the algebraic properties of the two coupling tensors $\B$ and $\Bc$, in particular,  the differences between the algebraic structures, namely symmetry, of these two tensors, as well as their definiteness, singularity and shape (in the sense defined below). The mutual influences between the coupling tensors and the compliance tensors $\Ac$ and $\Dc$ are eventually investigated. 

\section{Recall of the basic elements of the CLPT}
In the CLPT,  \cite{jones,vannucci_libro}, the mechanical behavior of a laminate is described by a constitutive law of the type
 \begin{equation}
 \label{eq:fundlaw}
 \left\{\begin{array}{c}\N \\\hline \M \end{array}\right\}=
 \left[\begin{array}{c|c}h\A & \frac{h^2}{2}\B \\\hline \frac{h^2}{2}\B & \frac{h^3}{12}\D\end{array}\right]
 \left\{\begin{array}{c}\beps \\\hline \bk\end{array}\right\},
 \end{equation}
 where $\N$ and $\M$ are respectively the extension and bending internal actions tensors, $\beps$ is the extension strain tensor of the mid-plane, $\bk$ the curvature tensor of the mid-plane, $h$ the laminate's thickness and, as previously said, $\A,\B$ and $\D$ are respectively the extension, coupling and bending stiffness tensors, homogeneous to an elasticity tensor and having the same classes of symmetry, i.e. the minor,
\be
\A_{ijkl}=\A_{jikl}=\A_{ijlk},
\ee
and major symmetries (of the indexes):
\be
\A_{ijkl}=\A_{klij}.
\ee
The major symmetries are actually the condition for a fourth-rank tensor to be defined as symmetric: $\A=\A^\top$, cf.  \cite{vannucci_alg}. Of course,  the same can be said also for $\B$ and $\D$. We adopt the Kelvin's formalism \cite{kelvin,kelvin1} for representing 2nd- and 4th-rank tensors, e.g.:
\be
\N=\left\{
\begin{array}{c}
N_1=N_{11}\\
N_2=N_{22}\\
N_6=\sqrt{2}N_{12}
\end{array}
\right\},\ \
\ee
and similarly for $\M,\beps$ and $\bk$, and
\be
\label{eq:kelvinmatrix}
\A\hspace{-1mm}=\hspace{-1mm}\left[
\begin{array}{ccc}
\A_{11}\hspace{-1mm}=\hspace{-1mm}\A_{1111}&\hspace{-1mm}\A_{12}\hspace{-1mm}=\hspace{-1mm}\A_{1122}&\hspace{-1mm}\A_{16}\hspace{-1mm}=\hspace{-1mm}\sqrt{2}\A_{1112}\\
\A_{12}\hspace{-1mm}=\hspace{-1mm}\A_{1122}&\hspace{-1mm}\A_{22}\hspace{-1mm}=\hspace{-1mm}\A_{2222}&\hspace{-1mm}\A_{26}\hspace{-1mm}=\hspace{-1mm}\sqrt{2}\A_{2212}\\
\A_{16}\hspace{-1mm}=\hspace{-1mm}\sqrt{2}\A_{1112}&\hspace{-1mm}\A_{26}\hspace{-1mm}=\hspace{-1mm}\sqrt{2}\A_{2212}&\hspace{-1mm}\A_{66}\hspace{-1mm}=\hspace{-1mm}2\A_{1212}
\end{array}
\right],
\ee
and the same for $\B,\D,\Ac,\Bc,\Dc$.

Tensors $\A,\B,\D$ are function of the layers' reduced stiffness $\Q_k$, orientation $\delta_k$ and stacking $z_k$;  for a $n-$plies laminate composed by different layers,  it is
 \begin{equation}
 \begin{split}
 \label{eq:ABDtens}
 &\A=\frac{1}{h}\sum_{k=1}^n({z_k}-{z_{k-1}})\Q_k(\delta_k),\\
& \B=\frac{1}{h^2}\sum_{k=1}^n({z_k^2}-{z_{k-1}^2})\Q_k(\delta_k),\\
 &\D=\frac{4}{h^3}\sum_{k=1}^n({z_k^3}-{z_{k-1}^3})\Q_k(\delta_k).
 \end{split}
 \end{equation}
For an equal-ply, i.e. composed of identical layers, laminate, the previous formulae reduce to
 \begin{equation}
  \label{eq:abcdeqply}
  \begin{split}
& \A=\sum_{k=1}^na_k\Q(\delta_k),\ 
 \B=\sum_{k=1}^nb_k\Q(\delta_k),\ \D=\sum_{k=1}^nd_k\Q(\delta_k),
\end{split}
 \end{equation}
where
\begin{equation}
\label{eq:coefABCD}
\begin{split}
&a_k=\frac{1}{n},\ b_k=\frac{1}{n^2}(2k-n-1),\\
&d_k=\frac{1}{n^3}\left[12k(k-n-1)+4+3n(n+2)\right].
\end{split}
\end{equation}
The converse of eq. (\ref{eq:fundlaw}) is
\begin{equation}
\label{eq:fundlawinv}
 \left\{\begin{array}{c}\beps \\\hline \bk\end{array}\right\}=
 \left[\begin{array}{c|c}\frac{1}{h}\mathcal{A} & \frac{2}{h^2}\mathcal{B} \\\hline \frac{2}{h^2}\mathcal{B}^\top & \frac{12}{h^3}\mathcal{D}\end{array}\right]
 \left\{\begin{array}{c}\gr{N} \\\hline \gr{M}\end{array}\right\},
 \end{equation}
 with, cf.  \cite{vannucci_libro},
  \begin{equation}
 \label{eq:inversetensorsABD}
 \begin{split}
 &\mathcal{A}=(\A-3\B\D^{-1}\B)^{-1},\\ 
 &\mathcal{B}=-3\mathcal{A}\B\D^{-1}=(-3\mathcal{D}\B\A^{-1})^\top=-3\A^{-1}\B\mathcal{D},\\
  &\mathcal{D}=(\D-3\B\A^{-1}\B)^{-1}.
\end{split}
 \end{equation}
Like $\A$ and $\D$, also $\Ac=\Ac^\top$ and $\Dc=\Dc^\top$, i.e. also $\Ac$ and $\Dc$ have the major symmetries, but not $\Bc$: generally speaking, $\Bc\neq\Bc^\top$. This is a fundamental difference between $\B$ and $\Bc$ and it is the object of one of the next Sections. It is also to be recalled  that normally $\A\neq\D$ and $\Ac\neq\Dc$: the extension and bending behaviors are different, generally speaking. When $\A=\D$ and $\B=\mathbb{O}$ we say that the laminate is quasi-homogeneous \cite{CompScTech01,vannucci01ijss}. We are interested here to a wider class of laminates, the {\it  Quasi-Homogeneous Coupled Laminates} (QHCL) \cite{kandil}, i.e. coupled laminates with $\A=\D$. In such a case, we will see, in general it is $\Ac\neq\Dc$.

\begin{figure}
\centering
\includegraphics[width=\columnwidth]{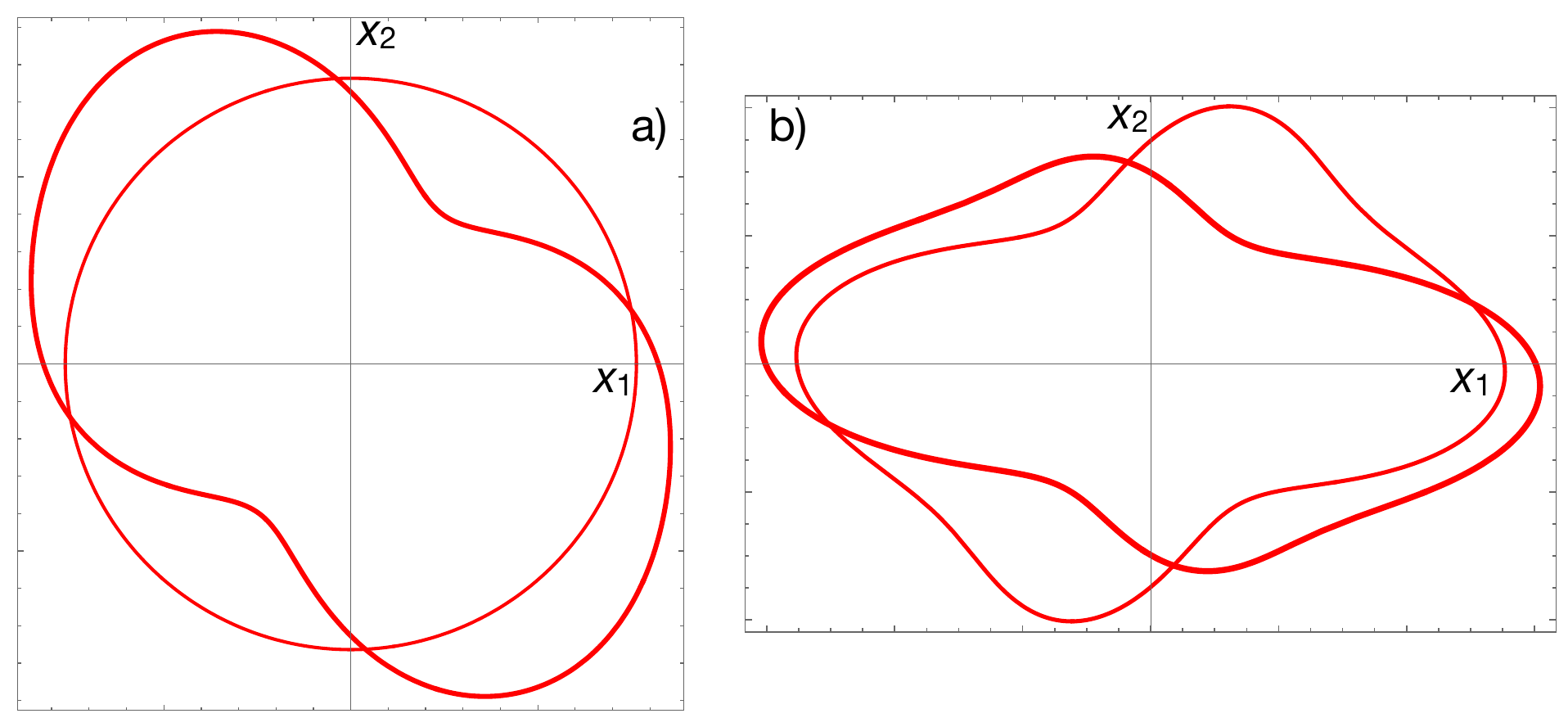}
\caption{An example of coupled laminate; a) $\A_{11}(\theta)$, thin, and $\D_{11}(\theta)$, thick, directional diagrams; b) the directional diagrams of the Young's modulus $E_A(\theta)$ for extension, thin, and $E_D(\theta)$ for bending, thick.}
\label{fig:1}
\end{figure}

It is worth  noting that the homogenization rules (\ref{eq:ABDtens}) or (\ref{eq:abcdeqply}) apply exclusively to the stiffness tensors, not to the compliance ones: the determination of $\Ac,\Bc$ and $\Dc$ can be done only through eq. (\ref{eq:inversetensorsABD}), which shows how much can be cumbersome to determine the compliance properties for coupled tensors. We remark also that $\Ac=\A^{-1}$ and $\Dc=\D^{-1}\iff\B=\mathbb{O}$. Only in this case, i.e. for uncoupled laminates, $\Ac$ is the inverse of $\A$ and $\Dc$ of $\D$ and, of course, $\Bc=\mathbb{O}$ too. 
That is why, for a coupled laminate, $\A$ and $\Ac$ have not the same material symmetries, generally speaking, and the same is true for $\D$ and $\Dc$. A simple example is the  laminate composed by identical orthotropic plies  whose stacking sequence is $[0^\circ,\ 60^\circ_{\ 2},\ 0^\circ,\ -60^\circ_{\ 2}]$. This laminate satisfies the Werren and Norris \cite{werren53} sufficient conditions for in-plane isotropy, but not for bending isotropy and it is coupled. As a consequence, $\A$ is isotropic, while $\B,\D,\Ac$ and $\Dc$ are completely anisotropic, as one can appreciate in the directional diagrams of Fig. \ref{fig:1} (recall that the Young's modulus is linked to $\Ac$ or $\Dc$ by the relations $E_A(\theta)=\frac{1}{\Ac_{11}(\theta)}, E_D(\theta)=\frac{1}{\Dc_{11}(\theta)})$; moreover, $\Bc\neq\Bc^\top$. The material of the layers for this example is carbon-epoxy T300-5208 \cite{TsaiHahn} whose elastic polar moduli are $T_0=26.88$ GPa, $T_1=24.74$ GPa, $R_0=19.71$ GPa, $R_1=21.43$ GPa, $\Phi_0=\Phi_1=0$, which gives, for the technical constants, $E_1=181$ GPa, $E_2=10.30$ GPa, $G_{12}=7.17$ GPA, $\nu_{12}=0.28$.


This is just a small example of the heterogeneity of the extension and bending behavior, on the one hand, and of the influence of coupling on the compliance behavior: $\A$ is isotropic but $\Ac$, and hence the engineering moduli like the Young's modulus $E$, are not. 

This example is also interesting to pointing out the difficulty of correctly defining the concept of material symmetries for a coupled laminate: $\A$ is isotropic, but $\Ac$ is completely anisotropic. So, it is questionable in a case like this one to affirm that the laminate is isotropic in extension. More appropriate is to consider just the symmetries of the elastic tensors, always possible to do and well identified by tensor invariants in the polar formalism, see the next Section.

\section{The CLPT by the polar formalism}
With the polar formalism, the Cartesian components in the Kelvin notation at a direction $\theta$ of a plane elastic tensor $\mathbb{T}$ are expressed as
\begin{equation}
\label{eq:mohr4}
{\begin{split}
&{\mathbb{T}_{11}{(}\theta{)}{=}{T}_{0}{+}{2}{T}_{1}{+}{R}_{0}\cos{4}\left({{{\varPhi}}_{0}{-}\theta}\right){+}{4}{R}_{1}\cos{2}\left({{{\varPhi}}_{1}{-}\theta}\right)},\\
&{\mathbb{T}_{12}{(}\theta{)}{=}{-}{T}_{0}{+}{2}{T}_{1}{-}{R}_{0}\cos{4}\left({{{\varPhi}}_{0}{-}\theta}\right)},\\
&\mathbb{T}_{16}(\theta)=\sqrt{2}\left[R_0\sin4\left(\varPhi_0-\theta\right)+2R_1\sin2\left(\varPhi_1-\theta\right)\right],\\
&{\mathbb{T}_{22}{(}\theta{)}{=}{T}_{0}{+}{2}{T}_{1}{+}{R}_{0}\cos{4}\left({{{\varPhi}}_{0}{-}\theta}\right){-}{4}{R}_{1}\cos{2}\left({{{\varPhi}}_{1}{-}\theta}\right)},\\
&{\mathbb{T}_{26}{(}\theta{)}{=}\sqrt{2}\left[{-}{R}_{0}\sin{4}\left({{{\varPhi}}_{0}{-}\theta}\right){+}{2}{R}_{1}\sin{2}\left({{{\varPhi}}_{1}{-}\theta}\right)\right]},\\
&{\mathbb{T}_{66}{(}\theta{)}{=}2\left[{T}_{0}{-}{R}_{0}\cos{4}\left({{{\varPhi}}_{0}{-}\theta}\right)\right]}.
\end{split}}
\end{equation}
What is important to remark is the fact that the moduli $T_0,T_1,R_0,R_1$ as well as the difference of the angles $\Phi_0-\Phi_1$ are tensor invariants; the value of one of the two polar angles, usually $\Phi_1$, fixes the frame. Moreover, all the elastic symmetries are determined by the following values of the invariants:
\begin{itemize}
\item ordinary orthotropy: $\Phi_0-\Phi_1=k\dfrac{\pi}{4}, \ k\in\{0,1\}$;
\item $R_0$-orthotropy: $R_0=0$,  \cite{vannucci02joe};
\item square symmetry: ${R_1=0}$;
\item isotropy: $R_0=R_1=0$.
\end{itemize}
It is immediate to see that $T_0$ and $T_1$ are the {\it isotropy invariants}, while $R_0,R_1$ and $\Phi_0-\Phi_1$ are the {\it anisotropy invariants}.
The above polar transformations apply to any tensor of the elastic type, hence  to  $\A,\B,\D,\Ac,\Dc$ too, {\it but not to} $\Bc$, because it is not symmetric (we consider this special tensor case below). 
We will denote by a superscript $A,B$ or $D$ a polar quantity of $\A,\B$ or $\D$ respectively, while a polar quantity of the basic layer has no superscript, and we will use capital letters for a polar quantity of the stiffness tensors and lower-case letters for the compliance tensors; so, e.g., $T_0^A,\Phi_1^A$ etc. indicate polar parameters of $\A$, while $t_0^A,\phi_1^A$ etc. the corresponding ones of $\Ac$, and similarly for $\B,\Bc,\D$ and $\Dc$.

The homogenization rules (\ref{eq:ABDtens}) apply to the polar components as well:
\begin{equation}
 \label{eq:compApolar}
 \A\ \ \rightarrow\ \
 \left\{
 \begin{split}
 &T_0^A=\frac{1}{h}\sum_{k=1}^n{T_0}_k(z_k-z_{k-1}),\\
 &T_1^A=\frac{1}{h}\sum_{k=1}^n{T_1}_k(z_k-z_{k-1}),\\
&R_0^Ae^{4i\Phi_0^A}=\frac{1}{h}\sum_{k=1}^n{R_0}_ke^{4i({\Phi_0}_k+\delta_k)}(z_k-z_{k-1}),\\
&R_1^Ae^{2i\Phi_1^A}=\frac{1}{h}\sum_{k=1}^n{R_1}_ke^{2i({\Phi_1}_k+\delta_k)}(z_k-z_{k-1}).
 \end{split}
 \right.
 \end{equation}
\begin{equation}
 \label{eq:compBpolar}
 \B\ \ \rightarrow\ \
 \left\{
 \begin{split}
 &T_0^B=\frac{1}{h^2}\sum_{k=1}^n{T_0}_k(z_k^2-z_{k-1}^2),\\
 &T_1^B=\frac{1}{h^2}\sum_{k=1}^n{T_1}_k(z_k^2-z_{k-1}^2),\\
&R_0^Be^{4i\Phi_0^B}=\frac{1}{h^2}\sum_{k=1}^n{R_0}_ke^{4i({\Phi_0}_k+\delta_k)}(z_k^2-z_{k-1}^2),\\
&R_1^Be^{2i\Phi_1^B}=\frac{1}{h^2}\sum_{k=1}^n{R_1}_ke^{2i({\Phi_1}_k+\delta_k)}(z_k^2-z_{k-1}^2).
 \end{split}
 \right.
 \end{equation}
\begin{equation}
 \label{eq:compDpolar}
 \D\ \ \rightarrow\ \
 \left\{
 \begin{split}
 &T_0^D=\frac{4}{h^3}\sum_{k=1}^n{T_0}_k(z_k^3-z_{k-1}^3),\\
 &T_1^D=\frac{4}{h^3}\sum_{k=1}^n{T_1}_k(z_k^3-z_{k-1}^3),\\
&R_0^De^{4i\Phi_0^D}=\frac{4}{h^3}\sum_{k=1}^n{R_0}_ke^{4i({\Phi_0}_k+\delta_k)}(z_k^3-z_{k-1}^3),\\
&R_1^De^{2i\Phi_1^D}=\frac{4}{h^3}\sum_{k=1}^n{R_1}_ke^{2i({\Phi_1}_k+\delta_k)}(z_k^3-z_{k-1}^3).
 \end{split}
 \right.
 \end{equation}
 We remark that, on the one hand,   the isotropic and anisotropic parts of all the tensors remain {\it separated} in the homogenization of the polar parameters, for all the tensors and, on the other hand, that {\it special orthotropies are preserved}: 
 \begin{equation}
 \begin{split}
 &{R_0}_k=0\ \forall k\ \Rightarrow\ R_0^A=R_0^B=R_0^D=0,\\
 &{R_1}_k=0\ \forall k\ \Rightarrow\ R_1^A=R_1^B=R_1^D=0.
 \end{split}
 \end{equation}
 So, if a laminate is composed, e.g., by layers that are all square symmetric, though different (it is the case of a laminate obtained using different layers, but all reinforced by balanced fabrics), then all the tensors $\A,\ \B$ and $\D$ will be square symmetric, regardless of the stacking sequence and of the ply orientations. The same is true for layers that are $R_0-$orthotropic, but {\it not for ordinarily orthotropic layers}: generally speaking, ordinary orthotropy is not preserved through the homogenization process. 

Particularly important is the case of laminates composed of identical plies; in such a case, eq. (\ref{eq:abcdeqply}) in the polar formalism gives
 \begin{equation}
  \label{eq:compApolareq}
\mathbb{A}\rightarrow\left\{
\begin{array}{l}
T_0^A=T_0,\\
T_1^A=T_1,\\
R_0^Ae^{4i\varPhi_0^A}={R_0e^{4i\varPhi_0}}{\left(\xi_1+i\xi_2\right)},\\
R_1^Ae^{2i\varPhi_1^A}={R_1e^{2i\varPhi_1}}{\left(\xi_3+i\xi_4\right)};
\end{array}\right.
\end{equation}
\begin{equation}
  \label{eq:compBpolareq}
\mathbb{B}\rightarrow\left\{
\begin{array}{l}
T_0^B=0,\\
T_1^B=0,\\
R_0^Be^{4i\varPhi_0^B}={R_0e^{4i\varPhi_0}}{\left(\xi_5+i\xi_6\right)},\\
R_1^Be^{2i\varPhi_1^B}={R_1e^{2i\varPhi_1}}{\left(\xi_7+i\xi_8\right)};
\end{array}\right.
\end{equation}
 \begin{equation}
   \label{eq:compDpolareq}
\hspace{10.2mm}\mathbb{D}\rightarrow\left\{
\begin{array}{l}
T_0^D=T_0,\\
T_1^D=T_1,\\
R_0^De^{4i\varPhi_0^D}={R_0e^{4i\varPhi_0}}(\xi_{9}+i\xi_{10}),\\
R_1^De^{2i\varPhi_1^D}={R_1e^{2i\varPhi_1}}(\xi_{11}+i\xi_{12}).
\end{array}\right.
\end{equation}
In the above equations, the quantities $\xi_i,i=1,...,12$ are the so-called {\it lamination parameters} \cite{TsaiHahn,vannucci_libro},   quantities accounting for the geometry of the stacking sequence (i.e. for orientations and position of the layers on the stack):
\begin{equation}
\label{eq:laminationparameters}
\begin{aligned}
&{{{{\xi}}_{1}{{+}}{i}{{\xi}}_{2}}{{=}}\mathop{\sum}\limits_{{k}{{=}}{1}}\limits^{n}a_k\hspace{0.33em}{{e}^{4i{{\delta}}_{k}}}},\hspace{1cm}
& &{{{{\xi}}_{3}{{+}}{i}{{\xi}}_{4}}{{=}}\mathop{\sum}\limits_{{k}{{=}}{1}}\limits^{n}a_k\hspace{0.33em}{{e}^{2i{{\delta}}_{k}}}},\\
&{{{{\xi}}_{5}{{+}}{i}{{\xi}}_{6}}{{=}}\mathop{\sum}\limits_{{k}{{=}}{1}}\limits^{n}{{b}_{k}\hspace{0.33em}{e}^{4i{{\delta}}_{k}}}},\hspace{1cm}
& &{{{{\xi}}_{7}{{+}}{i}{{\xi}}_{8}}{{=}}\mathop{\sum}\limits_{{k}{{=}}{1}}\limits^{n}{{b}_{k}\hspace{0.33em}{e}^{2i{{\delta}}_{k}}}},\\
&{{{{\xi}}_{9}{{+}}{i}{{\xi}}_{10}}{{=}}\mathop{\sum}\limits_{{k}{{=}}{1}}\limits^{n}{{d}_{k}\hspace{0.33em}{e}^{4i{{\delta}}_{k}}}},\hspace{1cm}
& &{{{{\xi}}_{11}{{+}}{i}{{\xi}}_{12}}{{=}}\mathop{\sum}\limits_{{k}{{=}}{1}}\limits^{n}{{d}_{k}\hspace{0.33em}{e}^{2i{{\delta}}_{k}}}},\\%
\end{aligned}
\end{equation}

So, through the polar formalism we can see that:
\begin{itemize}
\item the isotropic part of $\A$ and $\D$ is equal to that of the basic layer: $T_0^A=T_0^D=T_0,T_1^A=T_1^D=T_1$;
\item $\B$ is exclusively anisotropic, i.e. its isotropic part is null ($T_0^B=T_1^B=0$), so its average on $\{0,2\pi\}$  is zero.
\end{itemize}

The case of tensors $\B$ and $\Bc$ are special cases. In fact, for laminates composed of identical plies the condition $T_0^B=T_1^B$ gives that $\B$ is a {\it rari-constant tensor} \cite{vannucci2016}, i.e. a tensor having the Cauchy-Poisson symmetries \cite{pearson,love,benvenuto91} in addition to the minor and major ones (in other words, $\B$ is completely symmetric with respect to any permutation of the indices), which means that it is also
\be
\B_{1122}=\B_{1212}\ \rightarrow\ \B_{12}=\frac{\B_{66}}{2}.
\ee
To remark that for hybrid laminates, in general $T_0^B\neq T_1^B$, so $\B$ is not rari-constant for such laminates. 

About tensor $\Bc$, we have already remarked that it has not the major symmetries, i.e. $\Bc\neq \Bc^\top$. So, its situation is the opposite one of that of $\B$ for identical layers laminates: it has a lacking of tensor symmetries of the indexes. This special type of elastic tensor has been considered in the framework of the polar formalism \cite{vannucci10ijss}; in such a case, the tensor has nine independent  components:
\begin{equation}
\label{eq:bspecial}
\begin{split}
&\Bc_{11}(\theta)=t_0^B\hspace{-1mm}+\hspace{-1mm}2t_1^B\hspace{-1mm}+\hspace{-1mm}r_0^B\cos4(\phi_0^B\hspace{-1mm}-\hspace{-1mm}\theta)+2r_1^B\cos2(\phi_1^B\hspace{-1mm}-\hspace{-1mm}\theta)+\\
&\hspace{10mm}+2r_2^B\cos2(\phi_2^B-\theta),\\
&\Bc_{16}(\theta)=\sqrt{2}[-t_3^B+\hspace{-1mm}r_0^B\sin4(\phi_0^B\hspace{-1mm}-\hspace{-1mm}\theta)+\hspace{-1mm}2r_2^B\sin2(\phi_2^B-\hspace{-1mm}\theta)],\\
&\Bc_{12}(\theta)=-t_0^B\hspace{-1mm}+\hspace{-1mm}2t_1^B\hspace{-1mm}-\hspace{-1mm}r_0^B\cos4(\phi_0^B\hspace{-1mm}-\hspace{-1mm}\theta)\hspace{-1mm}+\hspace{-1mm}2r_1^B\cos2(\phi_1^B\hspace{-1mm}-\hspace{-1mm}\theta)-\\
&\hspace{10mm}-2r_2^B\cos2(\phi_2^B-\theta),\\
&\Bc_{61}(\theta)=\sqrt{2}[t_3^B+\hspace{-1mm}r_0^B\sin4(\phi_0^B-\hspace{-1mm}\theta)+2r_1^B\sin2(\phi_1^B-\hspace{-1mm}\theta)],\\
&\Bc_{66}(\theta)=2[t_0^B-r_0^B\cos4(\phi_0^B-\theta)],\\
&\Bc_{62}(\theta)=\sqrt{2}[-t_3^B\hspace{-1mm}-\hspace{-1mm}r_0^B\sin4(\phi_0^B\hspace{-1mm}-\hspace{-1mm}\theta)+\hspace{-1mm}2r_1^B\sin2(\phi_1^B-\theta)],\\
&\Bc_{21}(\theta)=-t_0^B\hspace{-1mm}+\hspace{-1mm}2t_1^B\hspace{-1mm}-\hspace{-1mm}r_0^B\cos4(\phi_0^B\hspace{-1mm}-\hspace{-1mm}\theta)\hspace{-1mm}-\hspace{-1mm}2r_1^B\cos2(\phi_1^B\hspace{-1mm}-\hspace{-1mm}\theta)+\\
&\hspace{10mm}+2r_2^B\cos2(\phi_2^B-\theta),\\
&\Bc_{26}(\theta)=\sqrt{2}[t_3^B-r_0^B\sin4(\phi_0^B-\hspace{-1mm}\theta)+2r_2^B\sin2(\phi_2^B-\hspace{-1mm}\theta)],\\
&\Bc_{22}(\theta)=t_0^B+2t_1^B+r_0^B\cos4(\phi_0^B\hspace{-1mm}-\hspace{-1mm}\theta)\hspace{-1mm}-\hspace{-1mm}2r_1^B\cos2(\phi_1^B\hspace{-1mm}-\hspace{-1mm}\theta)-\\
&\hspace{10mm}-2r_2^B\cos2(\phi_2^B-\theta).\\
\end{split}
\end{equation}These relations will be useful in the next Section for the analysis of tensor $\Bc$. To the usual polar components, in this case three more polar parameters appear: $t_3^B,r_2^B$ and $\phi_2^B$. 


To close this part, and resuming:
\begin{itemize}
\item $\B$ and $\Bc$ are  elastic tensors of a special type: for laminates composed of identical plies, $\B$ is rari-constant, while $\Bc$ is not, in general,  symmetric;
\item moreover, $\B$ and $\Bc$ are not definite, in the sense that, contrarily to $\A,\Ac,\D$ and $\Dc$, that are positive definite, $\B$ as well as $\Bc$ are not necessarily positive definite (this topic is considered below);
\item $\B$ and $\Bc$ can be singular;
\item $\B$ and $\Bc$ can have different shapes (in the sense defined below);
\item $\B$ and $\Bc$ can have different symmetries.
\end{itemize}

We examine all these aspects in the next Sections. 

\section{A decomposition of $\Bc$}
Because $\Bc\neq \Bc^\top$, we can always put
\be
\label{eq:decompB}
\Bc=\Bc^e+\Bc^c\ \rightarrow\ \Bc^c=\Bc-\Bc^e,
\ee
with $\Bc^e$ the symmetric, hence {\it elastic} in the classical sense, part of $\Bc$ and $\Bc^c$ the {\it complementary} part of $\Bc$ with respect to $\Bc^e$. So, because $\Bc^e={\Bc^e}^\top$, it is function only of the polar parameters $t_0^B,t_1^B,r_0^B,r_1^B,\phi_0^B$ and $\phi_1^B$ or, which is the same,
\be
\Bc^e=\Bc^e(t_0^B,t_1^B,t_3^B\hspace{-1mm}=\hspace{-1mm}0,r_0^B,r_1^B,r_2^B\hspace{-1mm}=\hspace{-1mm}r_1^B,\phi_0^B,\phi_1^B,\phi_2^B\hspace{-1mm}=\hspace{-1mm}\phi_1^B).
\ee
So, by eq. (\ref{eq:bspecial}), written for $\Bc$ and for $\Bc^e$, and (\ref{eq:decompB}), by subtraction we get
\be
\begin{split}
&\Bc^c_{11}=2[r_2^B\cos2(\phi_2^B-\theta)-r_1^B\cos2(\phi_1^B-\theta)],\\
&\Bc^c_{16}=-t_3^B+2[r_2^B\sin2(\phi_2^B-\theta)-r_1^B\sin2(\phi_1^B-\theta)],\\
&\Bc^c_{12}=-2[r_2^B\cos2(\phi_2^B-\theta)-r_1^B\cos2(\phi_1^B-\theta)],\\
&\Bc^c_{61}=t_3^B,\\
&\Bc^c_{66}=0,\\
&\Bc^c_{62}=-t_3^B,\\
&\Bc^c_{21}=-2[r_2^B\cos2(\phi_2^B-\theta)-r_1^B\cos2(\phi_1^B-\theta)],\\
&\Bc^c_{26}=t_3^B+2[r_2^B\sin2(\phi_2^B-\theta)-r_1^B\sin2(\phi_1^B-\theta)],\\
&\Bc^c_{22}=-2[r_2^B\cos2(\phi_2^B-\theta)-r_1^B\cos2(\phi_1^B-\theta)].
\end{split}
\ee
The un-symmetric part of $\Bc$ depends hence uniquely on the four invariants $t_3^B,r_1^B,r_2^B$ and $\phi_2^B-\phi_1^B$.

The question is: is it possible that $\Bc^c=\mathbb{O}$, i.e. that $\Bc=\Bc^\top$? In other words, can it happen, and when, that, like $\B$, also $\Bc\in \mathbb{E}$la, where $\mathbb{E}$la is the set of classical elastic tensors? This is considered in the next Section.

\section{Laminates with $\Bc=\Bc^\top$}
A first, partial solution to this problem has been given in \cite{vannucci13jota}; here, we give a complete solution in terms of stiffness moduli uniquely. To this end, from eq. (\ref{eq:inversetensorsABD})$_2$ we get that 
\be
\Bc=\Bc^\top\ \iff\ \Ac\B\D^{-1}=\Dc\B\A^{-1}.
\ee
To have an expression involving only stiffness tensors, we proceed as follows: right multiplying both terms of the last equation by $\A$ gives
\be
\Ac\B\D^{-1}\A=\Dc\B,
\ee
and, on one hand, by eq. (\ref{eq:inversetensorsABD})$_1$,
\be
\begin{split}
\Ac\B\D^{-1}\A&=(\A-3\B\D^{-1}\B)^{-1}\B\D^{-1}\A=\\
&=(\A-3\B\D^{-1}\B)^{-1}(\A^{-1}\D\B^{-1})^{-1}=\\
&=
\left[\A^{-1}\D\B^{-1}(\A-3\B\D^{-1}\B)\right]^{-1}=\\
&=\left(\A^{-1}\D\B^{-1}\A-3\A^{-1}\B\right)^{-1},
\end{split}
\ee
while, on the other hand, by eq. (\ref{eq:inversetensorsABD})$_3$,
\be
\begin{split}
\Dc\B&=(\D-3\B\A^{-1}\B)^{-1}\B=\\
&=(\D-3\B\A^{-1}\B)^{-1}\left(\B^{-1}\right)^{-1}=\\
&=\left[\B^{-1}(\D-3\B\A^{-1}\B)\right]^{-1}=\\
&=\left(\B^{-1}\D-3\A^{-1}\B\right)^{-1},
\end{split}
\ee
so that $\Bc=\Bc^\top\iff$
\be
\left(\A^{-1}\D\B^{-1}\A-3\A^{-1}\B\right)^{-1}=\left(\B^{-1}\D-3\A^{-1}\B\right)^{-1}.
\ee
Developing, we get successively
\be
\begin{split}
&\A^{-1}\D\B^{-1}\A-3\A^{-1}\B=\B^{-1}\D-3\A^{-1}\B\rightarrow\\
&\A^{-1}\D\B^{-1}\A=\B^{-1}\D
\end{split}
\ee
and finally the condition
\be
\label{eq:condsym}
\D\B^{-1}\A=\A\B^{-1}\D.
\ee
This same condition can be given in another, more useful form, introducing the tensor
\be
\H=\D\B^{-1}\A-\A\B^{-1}\D;
\ee
then
\be
\begin{split}
\H^\top&=\left(\D\B^{-1}\A-\A\B^{-1}\D\right)^\top=\\
&=\A^\top\B^{-\top}\D^\top-\D^\top\B^{-\top}\A^\top=\\
&=\A\B^{-1}\D-\D\B^{-1}\A=-\H,
\end{split}
\ee
which means that the matrix representing $\H$ in the Kelvin notation is skew. So, there are three alternative ways to write condition eq. (\ref{eq:condsym}), i.e. to impose that $\Bc=\Bc^\top$:
\begin{itemize}
\item $\D\B^{-1}\A\in \mathbb{E}\mathrm{la}$,
\item $\A\B^{-1}\D\in \mathbb{E}\mathrm{la}$,
\item $\H=\mathbb{O}$.
\end{itemize}
Taking the last of the above conditions, this reduces to only three scalar conditions for $\Bc=\Bc^\top$  :
\be
\H_{12}=0,\ \H_{13}=0,\ \H_{23}=0.
\ee
Through eqs. (\ref{eq:mohr4}) written for $\A,\B,\D$, these three conditions can be written using uniquely the stiffness polar components of these tensors. The computation is very cumbersome, but it can be carried out using a code for formal computation. The results, denoted  in the order conditions $C_1,C_2$ and $C_3$, given here only for the case of laminates made of identical plies (the results for the hybrid case are too much long and complicate to be transcribed) are
\be
\begin{split}
C_1&=
\frac{2}{R_0^B {R_1^B}^2\cos4\Phi_B} \left\{{R_0^B}^2  T_1 (R_1^A\cos2 \delta_A-\right.\\
&- R_1^D \cos2 \delta_D)+R_1^B \times\\
&\left[2R_0^A R_1^B R_1^D \sin4 (\delta_A+ \Phi_A)\sin2\delta_D+\right.\\
&+2R_0^B R_1^A R_1^D \sin2 (\delta_A-\delta_D)\sin4\Phi_B-\\
&-2R_0^D R_1^A R_1^B \sin2\delta_A\sin4(\delta_D+\Phi_D)+\\
&+R_0^B  T_1(R_0^D \cos4 (\delta_D+\Phi_D-\Phi_B)-\\
&\left.\left.-R_0^A \cos4 (\delta_A+\Phi_A-\Phi_B))\right]\right\}=0,
\end{split}
\ee
\be
\begin{split}
C_2&=
\frac{\sqrt{2}}{R_0^B {R_1^B}^2\cos4\Phi_B}\hspace{-1mm} \left\{\left(2 {R_1^B}^2 T_0-{R_0^B}^2 T_1\right)\times\right.\\
&\times(R_1^A\sin2 \delta_A-R_1^D\sin2 \delta_D)-\\
&+{R_0^B}^2 R_1^D [T_1 \sin2 \delta_D-R_1^A \sin2 (\delta_A-\delta_D)]+ \\
&+T_0{R_1^B}^2[R_0^A  \sin4(\delta_A+\Phi_A)-R_0^D \sin4 (\delta_D+\Phi_D)]+\\
&+2R_0^A {R_1^B}^2 R_1^D \sin2\delta_D\cos4(\delta_A+ \Phi_A)+\\
&+T_0 R_0^B R_1^B [R_1^A \sin2 (\delta_A\hspace{-1mm}-\hspace{-1mm}2 \Phi_B)\hspace{-1mm}-\hspace{-1mm}R_1^D  \sin2 (\delta_D\hspace{-1mm}-\hspace{-1mm}2 \Phi_B)]+\\
&+2R_0^B R_1^A R_1^B R_1^D \sin2 (\delta_A-\delta_D)\cos4\Phi_B-\\
&+T_1 R_0^B R_1^B[R_0^A  \sin4 (\delta_A+\Phi_A-\Phi_B)-\\
&-R_0^D  \sin4 (\delta_D+\Phi_D-\Phi_B)]-\\
&-R_0^A R_0^D {R_1^B}^2 \sin4 (\delta_A-\delta_D+\Phi_A-\Phi_D)-\\
&-2R_0^D R_1^A {R_1^B}^2 \sin2\delta_A\cos4(\delta_D+\Phi_D)+\\
&+R_0^B R_1^B [R_0^D R_1^A \sin2 (\delta_A-2( \delta_D+\Phi_D-\Phi_B))-\\
&\left.-R_0^A  R_1^D\sin2(\delta_D-2(\delta_A+ \Phi_A-\Phi_B))]\right\}=0,
\end{split}
\ee
\be
\begin{split}
C_3&=
\frac{\sqrt{2}}{R_0^B {R_1^B}^2\cos4 \Phi_B}\hspace{-1mm}\left\{R_1^A \left(2 {R_1^B}^2 T_0-{R_0^B}^2 T_1\right)\hspace{-1mm}\sin2 \delta_A+\right.\\
&+{R_0^B}^2 R_1^D [T_1 \sin2 \delta_D+R_1^A \sin2(\delta_A-\delta_D)]-\\
&-  T_0{R_1^B}^2 [2R_1^D \sin2 \delta_D+R_0^A  \sin4(\delta_A+\Phi_A)]+\\
&+2R_0^A {R_1^B}^2 R_1^D \sin2\delta_D\cos4( \delta_A+ \Phi_A)+\\
&+2R_0^B R_1^A R_1^B R_1^D \sin2 (\delta_A-\delta_D)\cos4 \Phi_B+\\
&+T_0 R_0^B R_1^B [R_1^D \sin2 (\delta_D-2 \Phi_B)-\\
&-R_1^A  \sin2 (\delta_A-2 \Phi_B)]+\\
&+ T_1 R_0^B R_1^B[R_0^A  \sin4 (\delta_A+\Phi_A-\Phi_B)-\\
&-R_0^D\sin4 (\delta_D+\Phi_D-\Phi_B)]+\\
&+R_0^D {R_1^B}^2 [T_0 \sin4 (\delta_D+\Phi_D)- \\
&-R_0^A  \sin4 (\delta_D+\Phi_D-\delta_A-\Phi_A)]-\\
&-2R_0^D R_1^A{R_1^B}^2 \sin2 \delta_A\cos4 (\delta_D+\Phi_D)+\\
&+R_0^B R_1^B[R_0^A R_1^D \sin2( \delta_D-2(\delta_A+ \Phi_A- \Phi_B))-\\
&\left.- R_0^D R_1^A  \sin2 (\delta_A-2 (\delta_D+\Phi_D-\Phi_B))]\right\}=0,
\end{split}
\ee
where $\Phi_A=\Phi_0^A-\Phi_1^A,\Phi_B=\Phi_0^B-\Phi_1^B$ and $\Phi_D=\Phi_0^D-\Phi_1^D$, three polar invariants of $\A,\B$ and $\D$ respectively, while $\delta_A=\Phi_1^A-\Phi_1^B$ and $\delta_D=\Phi_1^D-\Phi_1^B$ are the {\it shift angles with respect to} $\B$, in the order of $\A$ and $\D$. Depending only upon invariant quantities and shift angles, the above conditions are frame independent. 

Conditions $C_1,C_2,C_3$  are general but difficult to be analyzed; by a numerical optimization procedure like the one described in \cite{vannucci06,vannucci09algo,vincenti10}, it is possible to obtain laminates satisfying them and also other requirements, e.g. orthotropy of $\A$ or $\D$. More interesting is to consider such conditions in some particular cases, important for applications. Some of these cases are analyzed hereafter.

\subsection{Orthotropic co-axial stiffness}
The first particular case is that of a laminate with $\A,\B$ and $\D$ designed to be orthotropic and with the direction of the orthotropy axes of the three tensors that coincide, i.e.:
\be
\begin{split}
&\Phi_A=k_A\frac{\pi}{4},\ \Phi_B=k_B\frac{\pi}{4},\ \Phi_D=k_D\frac{\pi}{4},\\
&\delta_A=h_A\frac{\pi}{2},\ \delta_D=h_D\frac{\pi}{2},\\
& k_A,k_B,k_D,h_A,h_D\in\{0,1\}.
\end{split}
\ee
Then, condition $C_1$ becomes
\be
\label{eq:orthalig}
\frac{(-1)^{k_A}R_0^A-(-1)^{k_D}R_0^D}{(-1)^{h_ A}R_1^A-(-1)^{h_D}R_1^D}=\frac{(-1)^{k_B}R_0^B}{R_1^B},
\ee
which is a generalization of the condition already given in \cite{vannucci13jota}, where it was implicitly admitted that $h_A=h_D=0$, while conditions $C_2$ and $C_3$ are identically satisfied. It is worth to remark that the only condition of orthotropy is not sufficient to get a simple and unique condition for being $\Bc=\Bc^\top$: co-axiality of the orthotropy axes is also necessary. This mechanical condition is a strong one for simplifying the relations among $\A,\B$ and $\D$, as already shown in \cite{vannucci23}. To remark also the algebraic symmetry of eq. (\ref{eq:orthalig}), obtained thanks to the polar formalism and involving only tensor invariants or shift angles, through parameters $h_A$ and $h_D$.

\subsection{Extension isotropy and bending orthotropy}
\begin{figure}
\centering
\includegraphics[width=\columnwidth]{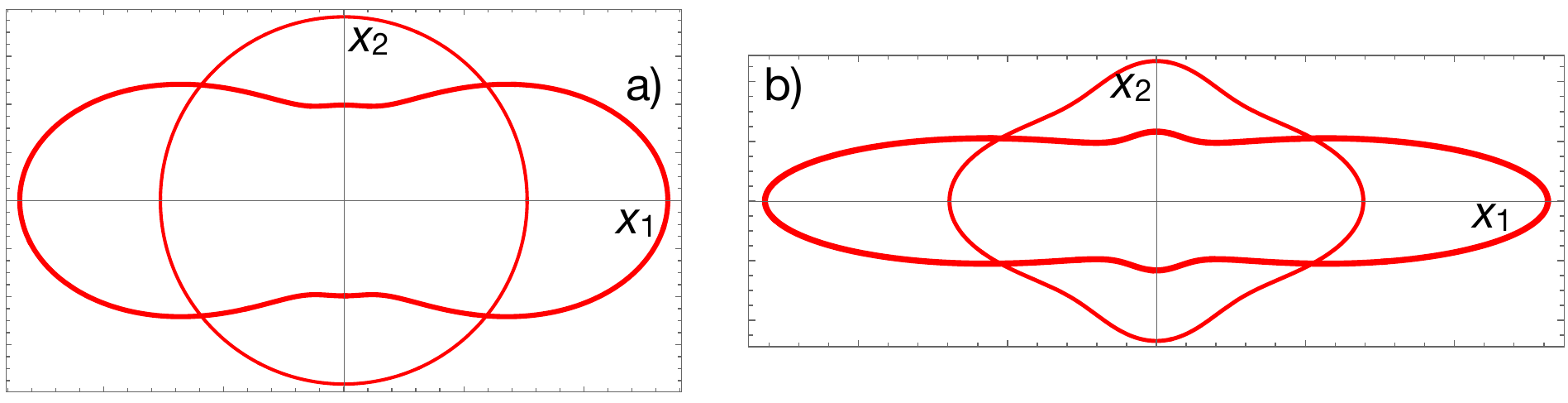}
\caption{An example of coupled laminate with $\A$ isotropic and $\B$ and $\D$ orthotropic; a) $\A_{11}(\theta)$, thin, and $\D_{11}(\theta)$, thick, directional diagrams; b) the directional diagrams of the Young's modulus $E_A(\theta)$ for extension, thin, and $E_D(\theta)$ for bending, thick. Material: carbon-epoxy T300-5202 \cite{TsaiHahn}.}
\label{fig:2}
\end{figure}

Let us consider now a laminate designed to have $\A$ isotropic and $\D$ orthotropic; then, once more $C_2$ and $C_3$ are identities, while $C_1$ becomes
\be
(-1)^{k_D}\frac{R_0^D}{R_1^D}\cos4(\delta_D-\Phi_B)=\frac{R_0^B}{R_1^B}\cos2\delta_D.
\ee
All the quantities concerning $\A$ disappear from $C_1$.
An example of this kind of laminates can be obtained simply changing the stacking sequence of the laminate considered previously to  $[0^\circ,\ 60^\circ_{\ 2},\ -60^\circ_{\ 2},\ 0^\circ]$. The directional diagrams of $\A_{11},\D_{11}$ and of the Young's moduli in extension and in bending are shown in Fig. \ref{fig:2}.

In the more particular case of $\B$ orthotropic and co-axial to $\D$,  the previous condition simplifies to
\be
(-1)^{k_D+h_D}\frac{R_0^D}{R_1^D}=(-1)^{k_B}\frac{R_0^B}{R_1^B}.
\ee

\subsection{Quasi-homogeneous coupled laminates}
Let us now consider the QHCL case, i.e. $\A=\D,\B\neq\mathbb{O}$. In this situation the three conditions $C_1,C_2$ and $C_3$ are identically satisfied: a sufficient condition for getting $\Bc=\Bc^\top$ is that $\A=\D$. This result can be shown directly, in another way: from eq. (\ref{eq:inversetensorsABD})$_{1,3}$ we get that,  if $\A=\D$,
\be
\label{eq:AcDc}
\Ac=(\A-3\B\D^{-1}\B)^{-1}=(\D-3\B\A^{-1}\B)^{-1}=\Dc,
\ee
so from eq. (\ref{eq:inversetensorsABD})$_2$
\be
 \Bc^\top=-3\Dc\B\A^{-1}=-3\Ac\B\D^{-1}=\Bc.
\ee
To remark that this result is valid also for hybrid laminates.

\subsection{Square-symmetric laminates}
We consider now laminates composed by square symmetric layers, i.e. layers all having $R_1=0\Rightarrow R_1^A=R_1^B=R_1^D=0$. In this case, $\Phi_1^A,\Phi_1^B$ and $\Phi_1^D$ are not defined, so $\Phi_A=\Phi_0^A,\Phi_B=\Phi_0^B,\Phi_D=\Phi_0^D$ and we redefine $\delta_A=\Phi_A-\Phi_B,\delta_D=\Phi_D-\Phi_B$. In the case of hybrid laminates, $C_1$ is an identity, while $C_2$ becomes
\be
\label{eq:sqsymlam1}
\begin{split}
&{R_0^A} {R_0^B} {T_0^D} \sin4\delta_A-{R_0^B} {R_0^D}{T_0^A} \sin4\delta_D=\\
&={R_0^A} {R_0^D} {T_0^B} \sin4({\delta_A}-{\delta_D}),
\end{split}
\ee
and $C_3=-C_2$. The case of laminates made of identical plies is more delicate, because in such a situation, $\B$ is singular, see below. Hence, $\H$ is not defined, nor $C_1,C_2,C_3$. However, it is still possible to calculate $\Bc$ through eqs. (\ref{eq:inversetensorsABD})$_{1,2}$ to get
\be
\label{eq:binvcomplet}
\Bc=-3(\A-3\B\D^{-1}\B)^{-1}\B\D^{-1},
\ee
and then impose $\Bc=\Bc^\top$, i.e. the three conditions
\be
\Bc_{12}-\Bc_{21}=0,\ \Bc_{13}-\Bc_{31}=0,\ \Bc_{23}-\Bc_{32}=0.
\ee
The first condition is an identity, while the second one is
\be
\begin{split}
\label{eq:elaBR1}
{R_0^A} \sin4\delta_ A={R_0^D} \sin4\delta_D,
\end{split}
\ee
as the third one. It is also apparent that this last condition can be obtained from eq. (\ref{eq:sqsymlam1}) putting $T_0^A=T_0^D,T_0^B=0$, as it is for identical plies.

\subsection{$R_0$-orthotropic laminates}
More exotic and intriguing is the case of $R_0$-orthotropic laminates \cite{vannucci02joe}, i.e. having $R_0^A=R_0^B=R_0^D=0$, a case that can be get when all the layers have $R_0=0$. A sufficient condition to have a ply with $R_0=0$ is that it is reinforced by fibers in equal quantities along two directions tilted of $45^\circ$. Now, $\Phi_0^A,\Phi_0^B$ and $\Phi_0^D$ are not defined, so $\Phi_A=\Phi_1^A,\Phi_B=\Phi_1^B,\Phi_D=\Phi_1^D$. Also in this case $\B$ is singular for identical plies, so proceeding like for the previous case, we get the three conditions
\be
\begin{split}
&{R_1^A} ({T_0^B} {T_1^D}-{T_0^D} {T_1^B}) \cos2
{\delta_ A}+\\
+&{R_1^D} ({T_0^A} {T_1^B}-{T_0^B} {T_1^A}) \cos2 {\delta_ D}=\\
=&{R_1^B} ({T_0^A} {T_1^D}-{T_0^D} {T_1^A}),
\end{split}
\ee
\be
\begin{split}
&\hspace{4mm}{R_1^A}\left[2 {R_1^B}^2 {T_0^D}-{R_1^B} {T_0^B} {T_0^D}+\right.\\
&\left.+{T_0^B} ({T_0^B} {T_1^D}-{T_0^D} {T_1^B})\right]
\sin2 {\delta_ A}+\\
&+{R_1^D}\left[{R_1^B} {T_0^A} {T_0^B}-2 {R_1^B}^2 {T_0^A}-\right.\\
&\left.-{T_0^B}(T_0^B {T_1^A}-{T_0^A} {T_1^B})\right]
\sin2 {\delta_ D}=\\
&={R_1^A} {R_1^D} {T_0^B} (2 {R_1^B}-{T_0^B}) \sin2 ({\delta_ A}-{\delta_ D}),
\end{split}
\ee
\be
\begin{split}
&\hspace{4mm}R_1^A \left[2 {R_1^B}^2 T_0^D+R_1^B T_0^B T_0^D+\right.\\
&\left.+T_0^B (T_0^B T_1^D-T_0^D T_1^B)\right]\sin2 \delta_A-\\
&-R_1^D\left[R_1^B T_0^A T_0^B+2 {R_1^B}^2 T_0^A+\right.\\
&\left.+T_0^B (T_0^B T_1^A-T_0^A T_1^B)\right] \sin2
\delta_D=\\
&=R_1^A R_1^DT_0^B (2 R_1^B+T_0^B) \sin2 (\delta_A-\delta_D).
\end{split}
\ee
For laminates made of identical plies, the first one of the above conditions becomes an identity while the two others coincide and are
\be
\label{eq:condbsymR0}
\begin{split}
&R_1^A\sin2\delta_A=R_1^D\sin2\delta_D.\\
\end{split}
\ee

\subsection{Hybrid isotropic coupled laminates}
A special case of coupled laminates is the one get by stacking together isotropic plies of different materials, like in the case of a bimetal. In this circumstance, $\A,\B$ and $\D$ depend uniquely on $T_0^A,T_1^A,T_0^B,T_1^B,T_0^D,T_1^D$. Using successively eqs. (\ref{eq:mohr4}), written for $\A,\B,\D$, and (\ref{eq:inversetensorsABD})$_{1,2}$ we obtain 
\be
\label{eq:bisotropic}
\Bc=\left[
\begin{array}{ccc}
t_0^B+2t_1^B&-t_0^B+2t_1^B&0\\
-t_0^B+2t_1^B&t_0^B+2t_1^B&0\\
0&0&2t_0^B
\end{array}
\right],
\ee
with
\be
\ \ 
t_0^B=\frac{3}{4}\frac{T_0^B}{3{T_0^B}^2-T_0^AT_0^D},\ t_1^B=\frac{3}{16}\frac{T_1^B}{3{T_1^B}^2-T_1^AT_1^D},
\ee
i.e. $\Bc$ is isotropic too and $\Bc=\Bc^\top$.

\section{Rari-constant $\Bc$}
In the previous section we have seen in which circumstances $\Bc\in\mathbb{E}$la. If the laminate is hybrid, in such a situation $\B$ and $\Bc$ belong hence to the same subspace of 4th-rank tensors, but if the plies are identical, this is no longer true, because $\B\in\mathbb{E}$la-rc, the set of rari-constant elastic tensors. So, the question is: can $\Bc\in\mathbb{E}$la-rc? The mathematical condition for this to happen \cite{vannucci2016} is
\be
t_0^B=t_1^B.
\ee
Unfortunately, the expression of this equation in terms of stiffness moduli, get using a code for formal computing, is too huge to be written and handled, in the general case. However, in some particular cases this can be done, e.g. for square symmetric laminates, i.e. $R_1^A=R_1^B=R_1^D=0$, when the last equation becomes particularly simple:
\be
R_0^A\cos4\delta_A+R_0^D\cos4\delta_D=0;
\ee
This  condition should be satisfied in addition to eq. (\ref{eq:elaBR1}), which happens only for particular values of $\delta_A$ and $\delta_D$, namely when
\be
\tan4\delta_A=-\tan4\delta_D\iff\delta_A+\delta_D=\frac{\pi}{4}.
\ee
Hence, in general for laminates of identical plies $\Bc\notin\mathbb{E}$la-rc.

\section{Definiteness of $\B$ and $\Bc$}
As previously said, $\B$ and $\Bc$ are not defined, in the sense that they do not define a positive, nor a negative, quadratic form. As a consequence, it is not possible to define bounds for the moduli of these two tensors independently from the two other ones, $\A$ and $\D$ \cite{vannucci23}. In particular, among the bounds defined for the polar invariants of  any elastic tensor \cite{vannucci_libro}
\be
\label{eq:bornes}
\begin{split}
&T_0-R_0>0,\\
&T_1(T_0^2-R_0^2)-2R_1^2[T_0-R_0\cos4(\Phi_0-\Phi_1)]>0,\\
&R_0\geq0,\\
&R_1\geq0,
\end{split}
\ee
only the last two are valid for $\B$ and $\Bc$, because actually $R_0$ and $R_1$ are the moduli of complex numbers, and as such intrinsically non negative quantities.  To notice that, because eq. (\ref{eq:bornes})$_{1,2}$ are no longer valid for $\B$ and $\Bc$, then $T_0^B,T_1^B,t_0^B,t_1^B$ are not necessarily positive, unlike for $\A,\D,\Ac,\Dc$.

Actually, the non definiteness of $\B$ can be seen directly: if  eq. (\ref{eq:ABDtens})$_2$ is written for a frame where the axis $z$ is changed to $\zeta$, oriented contrarily to $z$, then $\zeta_k=z_{n-k}\forall k=0,...,n$ and then we get, with the new axis,
\be
\begin{split}
 \B_\zeta&=\frac{1}{h^2}\sum_{k=1}^n({\zeta_k^2}-{\zeta_{k-1}^2})\Q_k(\delta_k)=\\
 &=\frac{1}{h^2}\sum_{k=1}^n({z_{n-k}^2}-{z_{n-k+1}^2})\Q_k(\delta_k)=\\
  &=-\frac{1}{h^2}\sum_{k=1}^n({z_{n-k+1}^2}-{z_{n-k}^2})\Q_k(\delta_k)=-\B_z,
 \end{split}
\ee
where $\B_\zeta$ is $\B$ written with the axis $\zeta$ and $\B_z$ with the axis $z$. So, inverting the axis makes change the sign of all the components of $\B$, which means that $\B$ cannot be defined. Using the same reasoning with eqs. (\ref{eq:compBpolar})$_{1,2}$, it can be easily checked that the same is true also for $T_0^B$ and $T_1^B$, of course in the case of hybrid laminates, because for the case of identical plies $T_0^B=T_1^B=0$. In this situation, $R_0^B$ and $R_1^B$ are not null, otherwise $\B=\mathbb{O}$, and necessarily positive, i.e. they are insensitive to a change of orientation of the laminate. So, the change of the sign of $\B$ is get by a change of the polar angles $\Phi_0$ and $\Phi_1$ in passing from the $z$ to the $\zeta$ frame, namely
\be
\label{eq:changeangles}
(\Phi_0^B)_\zeta=\frac{\pi}{4}+(\Phi_0^B)_z, \ (\Phi_1^B)_\zeta=\frac{\pi}{2}+(\Phi_1^B)_z.
\ee
The same considerations are valid for $\Bc$ too; this can be checked using eq. (\ref{eq:binvcomplet}):
\be
\begin{split}
\Bc_\zeta&=-3(\A-3\B_\zeta\D^{-1}\B_\zeta)^{-1}\B_\zeta\D^{-1}=\\
&=-3(\A-3(-\B_z)\D^{-1}(-\B_z))^{-1}(-\B_z)\D^{-1}=\\
&=3(\A-3\B_z\D^{-1}\B_z)^{-1}\B_z\D^{-1}=-\Bc_z.
\end{split}
\ee
Then, from eq. (\ref{eq:mohr4}) we get that in passing from $z$ to $\zeta$, $t_0^B$ and $t_1^B$ change of sign while $\phi_0^B$ and $\phi_1^B$ change like in eq. (\ref{eq:changeangles}).

\section{Singularity of $\B$ and $\Bc$}
In some cases $\B$ and $\Bc$ are singular, e.g., when $\B=\Bc=\mathbb{O}$. We investigate now when this happens. From eqs. (\ref{eq:kelvinmatrix}) and (\ref{eq:mohr4}), written for $\B$, we get 
\be
\begin{split}
\det\B&=16\left[T_1^B\left({T_0^B}^2-{R_0^B}^2\right)-\right.\\
&\left.-2{R_1^B}^2\left(T_0^B-R_0^B\cos4\Phi_B\right)\right].
\end{split}
\ee
Hence, the condition for $\B$ to be singular is
\be
T_1^B\left({T_0^B}^2-{R_0^B}^2\right)=2{R_1^B}^2\left(T_0^B-R_0^B\cos4\Phi_B\right).
\ee
For hybrid laminates, this is possible
\begin{itemize}
\item if $\B$ is orthotropic, see below, when
\be
T_1^B(T_0^B+(-1)^{k_B}R_0^B)=2{R_1^B}^2;
\ee
\item if the laminate is isotropic, like bimetal plates, when
\be
T_0^BT_1^B=0,
\ee
i.e. when only one among $T_0^B$ and $T_1^B$ is null (not both, which implies  $\B=\mathbb{O}$).
\end{itemize}

For laminates made of identical plies, $\B$ is singular $\iff$
\be
R_0^BR_1^B\cos4\Phi_B=0.
\ee 
This happens in the following cases:
\begin{itemize}
\item $\B$ is square symmetric: $R_1^B=0$;
\item $\B$ is $R_0$-orthotropic: $R_0^B=0$;
\item $\Phi_B=\dfrac{\pi}{8}+k\dfrac{\pi}{4}$.
\end{itemize}
In particular, because $\Phi_B$ is not a multiple of $\dfrac{\pi}{4}$, an orthotropic laminate of identical plies cannot have $\B$ singular.

Using the theorem of Binet in eq. (\ref{eq:inversetensorsABD})$_2$ gives
\be
\det\Bc=-27\frac{\det\Ac\det\B}{\det\D};
\ee
because $\Ac$ and $\D$ are nonsingular,
\be
\det\Bc=0\iff\det\B=0.
\ee

\section{Shapes of $\B$ and $\Bc$}
We call here, for the sake of shortness, {\it shape of } $\B$ the way the matrix representing $\B$ is filled in, e.g. when it is orthotropic and in which axes. This is of some importance in applications, when some coupling effect is to be designed.

We consider just the case of laminates made of identical plies, because the hybrid case does not allow to obtain general results, unless the simple case of isotropic coupled laminates, already considered before (the matrix representing $\B$ is like the one of eq. (\ref{eq:bisotropic}), it is sufficient to replace $t_0^B,t_1^B$ by $T_0^B,T_1^B$).
Considering the basic layer to be orthotropic, like it  always is in reality, from eqs. (\ref{eq:compBpolareq}) and (\ref{eq:mohr4}), written for $\B$ and $\theta=\Phi_1^B$, making equal the two expressions of $\B_{11}$ and of $\B_{22}$ we obtain the two equations
\be
\left\{
\begin{array}{ll}
(-1)^kR_0\xi_5+4R_1\xi_7=&R_0^B\cos4\Phi_B+4R_1^B,\medskip\\
(-1)^kR_0\xi_5-4R_1\xi_7=&R_0^B\cos4\Phi_B-4R_1^B;
\end{array}
\right.
\ee
summing and subtracting the second from the first one, gives
\be
\label{eq:lampar1}
\xi_5=(-1)^k\frac{R_0^B}{R_0}\cos4\Phi_B,\ \ \xi_7=\frac{R_1^B}{R_1}.
\ee
Making the same with components $\B_{16}$ and  $\B_{26}$ gives the equations
\be
\left\{
\begin{array}{rr}
(-1)^kR_0\xi_6+2R_1\xi_8=&R_0^B\sin4\Phi_B,\medskip\\
-(-1)^kR_0\xi_6+2R_1\xi_8=&-R_0^B\sin4\Phi_B;
\end{array}
\right.
\ee
and again summing and subtracting we get
\be
\label{eq:lampar2}
\xi_6=(-1)^k\frac{R_0^B}{R_0}\sin4\Phi_B,\ \ \xi_8=0.
\ee
Finally, in the most general case the matrix representing $\B$ is 
\be
\B\hspace{-0.7mm}=\hspace{-0.7mm}(-1)^k\hspace{-0.7mm}R_0\hspace{-0.7mm}\left[
\begin{array}{ccc}
\hspace{-0.7mm}\xi_5\hspace{-0.7mm}+\hspace{-0.7mm}4(-1)^k\rho\xi_7&\hspace{-0.7mm}-\xi_5&\hspace{-0.7mm}\sqrt{2}\xi_6\\
\hspace{-0.7mm}-\xi_5&\hspace{-0.7mm}\xi_5\hspace{-0.7mm}-\hspace{-0.7mm}4(-1)^k\rho\xi_7&\hspace{-0.7mm}-\sqrt{2}\xi_6\\
\hspace{-0.7mm}\sqrt{2}\xi_6&\hspace{-0.7mm}-\sqrt{2}\xi_6&\hspace{-0.7mm}-2\xi_5
\end{array}\hspace{-0.7mm}
\right]\hspace{-0.7mm},
\ee
with $\rho$ the {\it anisotropy ratio} of the basic layer:
\be
\rho=\frac{R_1}{R_0}.
\ee
We remark that it is sufficient to know three lamination parameters to determine entirely $\B$. It is also worth noting that eqs. (\ref{eq:lampar1}) and (\ref{eq:lampar2}) give  the lamination parameters as functions of only invariant mechanical parameters of the basic layer and of $\B$. This is  true also for the other lamination parameters concerning $\A$ and $\D$ and gives the relation between the geometry of the stack and the mechanical properties of the layer and of the laminate.

Let us now consider some usual  cases of laminates.

\subsection{Angle-ply coupled laminates}
In angle-ply laminates, the layers are oriented, at equal number, at $\pm\delta$. Then, because $\sum_{k=1}^nb_k=0$, see \cite{vannucci_libro},
\be
\label{eq:angpxi}
\begin{split}
&\xi_5=\sum_{k=1}^nb_k\cos4\delta=0,\\
&\xi_6=\sum_{k=1}^nb_k\sin4(\pm\delta)=\sin4\delta(\sum_kb_k^+-\sum_kb_k^-)=\\
&\hspace{3mm}=2\sin4\delta\sum_kb_k^+=-2\sin4\delta\sum_kb_k^-,\\
&\xi_7=\sum_{k=1}^nb_k\cos2\delta=0,
\end{split}
\ee
where, for the sake of shortness, $b_k^+$ are the coefficient of the plies with orientation $+\delta$ and $b_k^-$ those with orientation $-\delta$. As a consequence, 
\be
\label{eq:angleply}
\B=2\sqrt{2}(-1)^k\sum_kb_k^+R_0\sin4\delta
\left[
\begin{array}{rrr}0&0&1\\0&0&-1\\1&-1&0\end{array}
\right].
\ee
By consequence, using the converse of eq. (\ref{eq:mohr4}), see \cite{vannucci_libro}, we get that
\be
R_0^B=2(-1)^k\sum_kb_k^+R_0\sin4\delta,\  R_1^B=0,\ \Phi_0^B=\frac{\pi}{8}.
\ee
To remark that in such a case, $\B$ is singular and, generally speaking,  $\Bc\neq\Bc^\top$ and $\Bc$ has not the same shape of $\B$, as it can be proved  using  the last expression of $\B$ into eq. (\ref{eq:binvcomplet}) (the expression of $\Bc$ is not given as too long to be written). However, if $\A$ and $\D$ are orthotropic and coaxial, then $\Bc$ will have the same shape of $\B$ but it will remain un-symmetric:
\be
\label{eq:casostranoA}
\begin{split}
\Bc(\theta)=\left[\begin{array}{ccc}
0&0&\Bc_{16}(\theta)\\0&0&\Bc_{26}(\theta)\\\Bc_{61}(\theta)&\Bc_{62}(\theta)&0
\end{array}
\right];
\end{split}
\ee 
Because $\Bc\neq\Bc^\top$,  it cannot be represented by the  polar formalism for a tensor of $\mathbb{E}$la, eq. (\ref{eq:mohr4}) and   the converse of eq. (\ref{eq:bspecial}) must be used to get the polar parameters of $\Bc$ when it is not symmetric, cf. \cite{vannucci10ijss}. The general case gives very long expressions, so we   consider the particular and simpler case of $\delta_A=\delta_D=0,k_A=k_D=0$, for which we give the final result:
\be
\begin{split}
&\Bc_{16}={\frac{3 ({R_1^A}-{T_1}) \B_{16} }{4 ({T_0}-{R_0^D}) \left[({R_0^A}+{T_0}) {T_1}-2 {R_1^A}^2\right]-6{T_1} \B_{16} ^2}},\\
&\Bc_{26}={\frac{3 ({R_1^A}+{T_1}) \B_{16} }{4 ({T_0}-{R_0^D}) \left[({R_0^A}+{T_0}) {T_1}-2 {R_1^A}^2\right]-6 {T_1} \B_{16}^2}},\\
&\Bc_{61}={\frac{3 ({R_1^D}-{T_1}) \B_{16} }{4 ({T_0}-{R_0^A}) \left[({R_0^D}+{T_0}) {T_1}-2 {R_1^D}^2\right]-6 {T_1} \B_{16}^2}},\\
&\Bc_{62}={\frac{3 ({R_1^D}+{T_1}) \B_{16} }{4 ({T_0}-{R_0^A}) \left[({R_0^D}+{T_0}) {T_1}-2 {R_1^D}^2\right]-6 {T_1} \B_{16}^2}}.
\end{split}
\ee

It is worth noting that conditions (\ref{eq:angpxi})$_{1,3}$ are  satisfied by two different kind of special orthotropies of $\B$:
\begin{itemize}
\item $\B\ R_0-$ orthotropic: $R_0^B=0$,
\item $\B$ square-symmetric: $R_1^B=0$,
\end{itemize}
but also by a completely anisotropic  $\B$:
\be
\cos4\Phi_B=0\Rightarrow\Phi_B=\frac{\pi}{8}.
\ee
We notice also that the angle-ply stack is not the only way to obtain $\B$ of the same shape as in eq. (\ref{eq:angleply}); the necessary and sufficient condition is that $\xi_5=\xi_7=0$, which can be obtained also by other stacking sequences.

\subsection{Cross-ply coupled laminates}
In this case, $\delta_k\in\{0,\frac{\pi}{2}\}\ \forall k=1,...,n.$ As a consequence, $\A$ and $\D$ are automatically orthotropic and
\be
\begin{split}
&\xi_5=\sum_{k=1}^nb_k\cos4\delta_k=\sum_{k=1}^nb_k=0,\\
&\xi_6=\sum_{k=1}^nb_k\sin4\delta_k=0,\\
&\xi_7=\sum_{k=1}^nb_k\cos2\delta_k=\sum_kb_k\cos0+\sum_jb_j\cos\pi=\\
&\hspace{3mm}=2\sum_kb_k=-2\sum_jb_j,
\end{split}
\ee
where the sum over $k$ is for plies at $0^\circ$ and that over $j$ for those at $90^\circ$. 
A quick glance at eqs. (\ref{eq:lampar1})$_1$ and (\ref{eq:lampar2})$_1$ shows that $\xi_5=\xi_6=0\iff R_0^B=0$: the coupling of a cross-ply laminate is necessarily $R_0$-orthotropic. 
Finally
\be
\begin{split}
\B&=4(-1)^kR_0\left[\begin{array}{ccc}(-1)^k\rho\xi_7&0&0\\0&-(-1)^k\rho\xi_7&0\\0&0&0\end{array}\right]=\\
&=8R_1\sum_kb_k\left[\begin{array}{ccc}1&0&0\\0&1&0\\0&0&0\end{array}\right].
\end{split}
\ee
$\B$ is of course singular. Proceeding for $\Bc$ like for angle-ply laminates, we find again that $\Bc\neq\Bc^\top$ but, unlike angle-plies, $\Bc$ does not have, in general, the same shape of $\B$:
\be
\Bc=\left[\begin{array}{ccc}\Bc_{11}&\Bc_{12}&0\\\Bc_{21}&\Bc_{22}&0\\0&0&0\end{array}\right];
\ee
the expression of the $\Bc_{ij}$s is not given here as too much long.

\subsection{Specially orthotropic laminates}
If the basic layer has $R_1=0$, then $R_1^A=R_1^B=R_1^D=0$, $\xi_7$ and $\xi_8$ are not defined, $\B $ and $\Bc$ are  singular, as it was already found, and $\Bc\neq\Bc^\top$. The expression of $\Bc$ is omitted as too much long.

If now $R_0=0$ for the basic layer, then $R_0^A=R_0^B=R_0^D=0$  and now $\xi_5$ and $\xi_6$ are not defined. The matrix of $\B$ reduces to
\be
\B=4R_1\xi_7\left[\begin{array}{ccc}1&0&0\\0&-1&0\\0&0&0\end{array}\right],
\ee
confirming the predicted result that $\B$ is singular, like $\Bc$, which in general will not be symmetric. However, if $\delta_A=\delta_D=0$, i.e., if $\A$ and $\D$ are coaxial with $\B$, then 
\be
\Bc=\left[\begin{array}{ccc}\Bc_{11}&\Bc_{12}&0\\\Bc_{12}&\Bc_{22}&0\\0&0&0\end{array}\right],
\ee
i.e. $\Bc=\Bc^\top$ (the $\Bc_{ij}$s are omitted because too much long).

\section{Influence of $\B$ on the compliances $\Ac$ and $\Dc$}
As  introduced, and shown in the two previous examples, the existence of a coupling modifies the compliances of the laminate, in the sense that the elastic symmetries are in general not preserved, e.g. $\Ac$ can be anisotropic or orthotropic also when $\A$ is isotropic) nor the shape (e.g. $\A$ and $\Ac$ can be both orthotropic but with the orthotropy axes shifted), the same for $\D$ and $\Dc$. Hence, the question is: can  the shape and symmetries of $\Ac$ and $\Dc$ be predicted knowing $\A$ and $\D$ when $\B\neq\mathbb{O}$?

Unfortunately, it is not possible to have a general response to this question: on the one hand, the situations can be very different and impossible to be reduced to a unique set of conditions, on the other hand, equations are so complicate that they do not allow to extract some useful and simple results. However,  some interesting cases can be analyzed, they are detailed hereafter.

\subsection{Isotropic QHCL}
Let us consider first the case of coupled laminates composed by identical layers, designed to have $\A=\D$ and isotropic: $R_0^A=R_0^D=R_1^A=R_1^D=0$. The question is: how are tensors $\Ac$ and $\Dc$? First of all, some examples of this case are the following stacking sequences of QHCL : 
\be
\begin{split}
&[\alpha,  \alpha,  \beta,  \alpha,  \beta,  \gamma,  \beta,  \alpha,  \alpha,  \alpha,  \gamma,  \beta,  \gamma,  \gamma,  \beta,  \gamma,  \beta,  \gamma],\\
&[\alpha,  \alpha,  \beta,  \alpha,  \gamma,  \beta,  \beta,  \alpha,  \alpha,  \alpha,  \beta,  \gamma,  \gamma,  \gamma,  \gamma,  \beta,  \beta,  \gamma],\\
&[\alpha,  \alpha,  \beta,  \alpha,  \gamma,  \beta,  \gamma,  \alpha,  \alpha,  \alpha,  \beta,  \beta,  \gamma,  \gamma,  \gamma,  \beta,  \beta,  \gamma],\\
&[\alpha,  \beta,  \gamma,  \gamma,  \beta,  \beta,  \gamma,  \alpha,  \beta,  \beta,  \alpha,  \gamma,  \alpha,  \gamma,  \alpha,  \alpha,  \gamma,  \beta],\\
&[\alpha,  \beta,  \gamma,  \gamma,  \beta,  \gamma,  \beta,  \gamma,  \alpha,  \alpha,  \gamma,  \beta,  \alpha,  \alpha,  \beta,  \beta,  \alpha,  \gamma].\\
\end{split}
\ee
When the identical layers are at least orthotropic and  $\alpha,\beta,\gamma$ are either $0^\circ,60^\circ$ or $-60^\circ$, then all of them have the same $\A,\Ac,\D$ and $\Dc$  but  different $\B$ and $\Bc$. Moreover, $\A=\D,\Ac=\Dc$ and isotropic; eq. (\ref{eq:condbsymR0}) is satisfied, so $\Bc$ is symmetric for all the cases.

 Using eqs. (\ref{eq:inversetensorsABD}) and (\ref{eq:mohr4}) written for $\theta=\Phi_1^B$ in this case, we can compute the expression of $r_0^A$ and $r_1^A$, that are equal to $r_0^D$ and $r_1^D$ respectively, thanks to eq. (\ref{eq:AcDc}). These quantities are given by the reverse of eq. (\ref{eq:mohr4}), see \cite{vannucci_libro}:
\be
\begin{split}
&r_0^A=\frac{1}{8}\sqrt{(\Ac_{11}-2\Ac_{12}-2\Ac_{66}+\Ac_{22})^2+8(\Ac_{16}-\Ac_{26})^2},\\
&r_1^A=\frac{1}{8}\sqrt{(\Ac_{11}-\Ac_{22})^2+2(\Ac_{16}+\Ac_{26})^2}.
\end{split}
\ee
Developing the computations gives the following two conditions:
\be
\label{eq:condpresiso}
\begin{split}
r_0^A=0\iff &{R_1^B} \left[36 {R_1^B}^4 {T_0}^2-12 {R_1^B}^2 {T_0}^3 {T_1}+\right.\\
&+\left(9 {R_0^B}^4+{T_0}^4\right) {T_1}^2+\\
&\left.+6{R_0^B}^2 {T_0} {T_1} \hspace{-1mm}\left({T_0} {T_1}-6 {R_1^B}^2\right) \cos8 {\Phi_B}\right]\hspace{-1mm}=0,\\
r_1^A=0\iff&{R_0^B} {R_1^B} \left[18 {R_1^B}^4 {T_0}^2+\left({T_0}^2-3{R_0^B}^2\right)^2 {T_1}^2-\right.\\
&-6 {R_1^B}^2 {T_0} {T_1}\left({T_0}^2 \hspace{-1mm}-3 {R_0^B}^2\right)(1 \hspace{-1mm}-\cos8 {\Phi_B}) -\\
&\left.-18 {R_1^B}^4 {T_0}^2 \cos8 {\Phi_B}\right]=0
\end{split}
\ee
\subsubsection{Case  $R_1^B=0$.} 
\begin{figure}
\centering
\includegraphics[width=.7\textwidth]{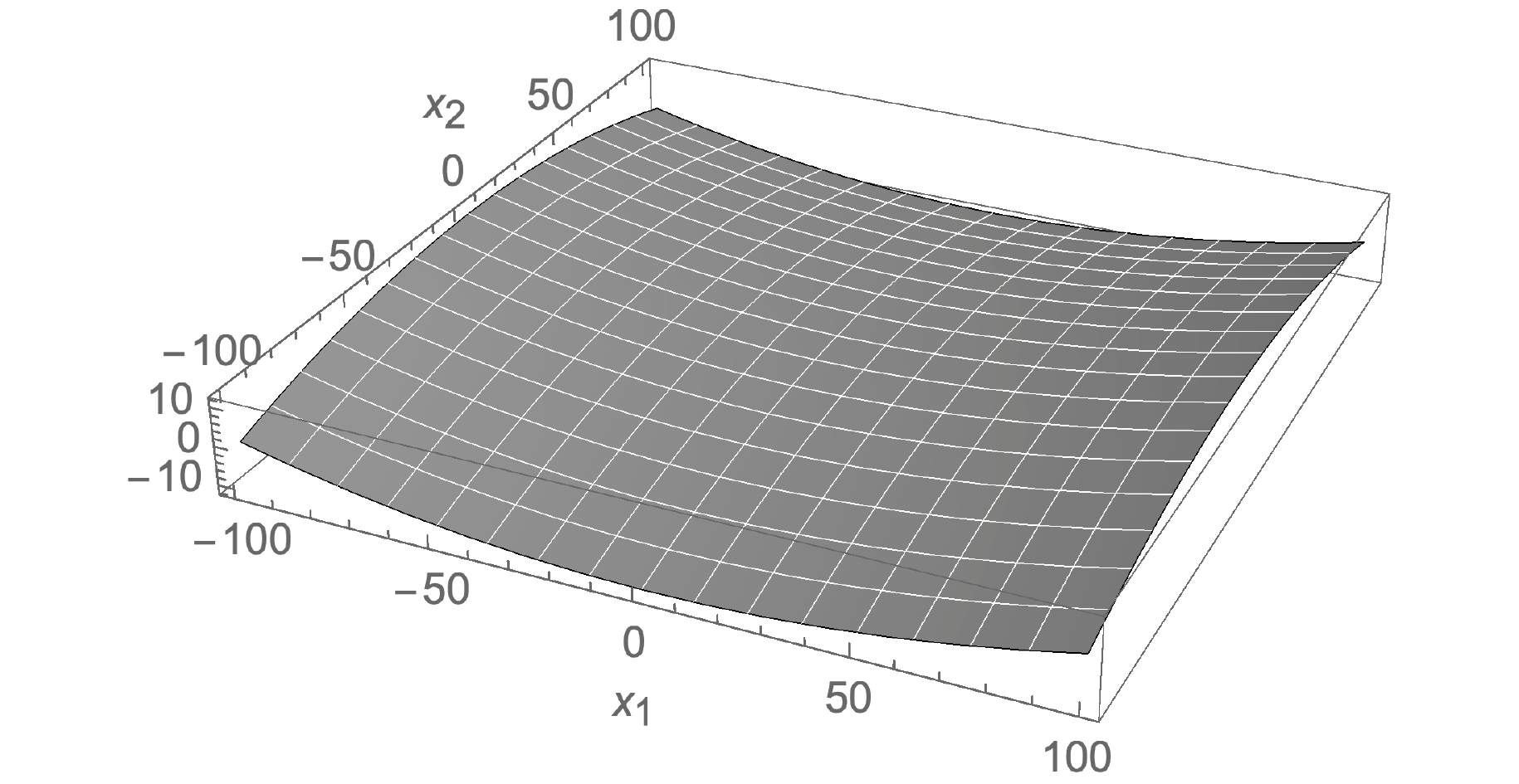}
\caption{An example of a coupled laminate with $\Ac=\Dc$ and isotropic, made of layers with $R_1=0$ and submitted to in-plane stretch: the deformed surface is a minimal surface (units: mm).}
\label{fig:3}
\end{figure}

This is a sufficient condition for preserving the isotropy of $\Ac$ and $\Dc$.  We  recall that if the basic layer is square symmetric,   automatically  $R_1^A=R_1^B=R_1^D=0$ (but this is not necessary to obtain $R_1^B=0$). 
In such a situation, we get
\be
\label{eq:r0strano}
\begin{split}
&t_0^A=t_0^D=\frac{T_0}{4(T_0^2-3{R_0^B}^2)}, \ t_1^A=t_1^D=\frac{1}{16T_1}\\
&t_0^B=t_1^B=r_1^B=0,\  r_0^B=\frac{3R_0^B}{4(T_0^2-3{R_0^B}^2)},\ \phi_0^B=\Phi_0^B,
\end{split}
\ee
i.e., $\Bc$ is square symmetric, like $\B$. Moreover, $\forall \theta$
\be
\Ac=\left[\begin{array}{ccc}
2t_1^A+t_0^A&2t_1^A-t_0^A&0\\2t_1^A-t_0^A&2t_1^A+t_0^A&0\\0&0&2t_0^A
\end{array}
\right],\ \ \Dc=\Ac
\ee
and
\be
\Bc(\theta)=r_0^B\left[\begin{array}{ccc}
\cos4\theta&-\cos4\theta&-\sqrt{2}\sin4\theta\\-\cos4\theta&\cos4\theta&\sqrt{2}\sin4\theta\\-\sqrt{2}\sin4\theta&\sqrt{2}\sin4\theta&-2\cos4\theta
\end{array}
\right].
\ee
So, if $\theta=0$, 
\be
\Bc(\theta=0)=r_0^B\left[\begin{array}{ccc}
1&-1&0\\-1&1&0\\0&0&-2
\end{array}
\right];
\ee
let us consider such a plate acted upon only by in-plane forces $\N=(N_1,N_2,N_6)$, a situation of interest in practice and considered also for other cases in the following; by eq. (\ref{eq:fundlawinv}) we get
\be
\label{eq:curvaturestrano}
\kappa_1=-\kappa_2=\frac{2}{h^2}r_0^B(N_1-N_2),\ \kappa_6=-\frac{4}{h^2}r_0^BN_6.
\ee
Because the effect of $h$ on the curvatures varies like $\frac{1}{h^2}$, the deformation of thin plates can be very important.
If $N_1=N_2,N_6=0$, the plate will not bend,  nor twist, if $N_2=-N_1$, or $N_1=0, N_2\neq0$ or $N_1\neq0,N_2=0$, and $N_6=0$, then $\kappa_1=-\kappa_2$ everywhere: the Gaussian curvature $K=\kappa_1\kappa_2$ is  negative, so the deformed surface is made of hyperbolic points. Moreover, the plate bends as a minimal surface, because the mean curvature $H=\frac{1}{2}(\kappa_1+\kappa_2)$ is null everywhere \cite{vannucci_alg,pressley}. As an example, consider a simple square plate made of two  layers with $R_1=0$ with the sequence $[-22.5^\circ,22.5^\circ]$. The material is a carbon-epoxy balanced fabric \cite{gay14} with the following characteristics: $E_1=E_2=5.4\times10^4$ MPa, $G_{12}=4\times10^3$ MPa, $\nu_{12}=0.045$, thickness $h_L=0.16$ mm. The plate has dimensions $200\times200$ mm and is acted upon by the in-plane forces $N_1=2$ N/mm, $N_2=-2$ N/mm: tension along the $x_1$ axis and compression along the $x_2$ axis.
With these data, we get $h=0.32$ mm, $T_0=1.49\times10^4$ MPa and $R_0^B=5.45\times10^3$ MPa, that gives, according to eqs. (\ref{eq:r0strano}) and (\ref{eq:curvaturestrano}), $r_0^B=3.07\times10^{-5}$ MPa$^{-1}$ and $\kappa_1=1.2\times10^{-2}$ mm$^{-1}$. The result is shown in Fig. \ref{fig:3}.

\begin{figure}
\centering
\includegraphics[width=.7\textwidth]{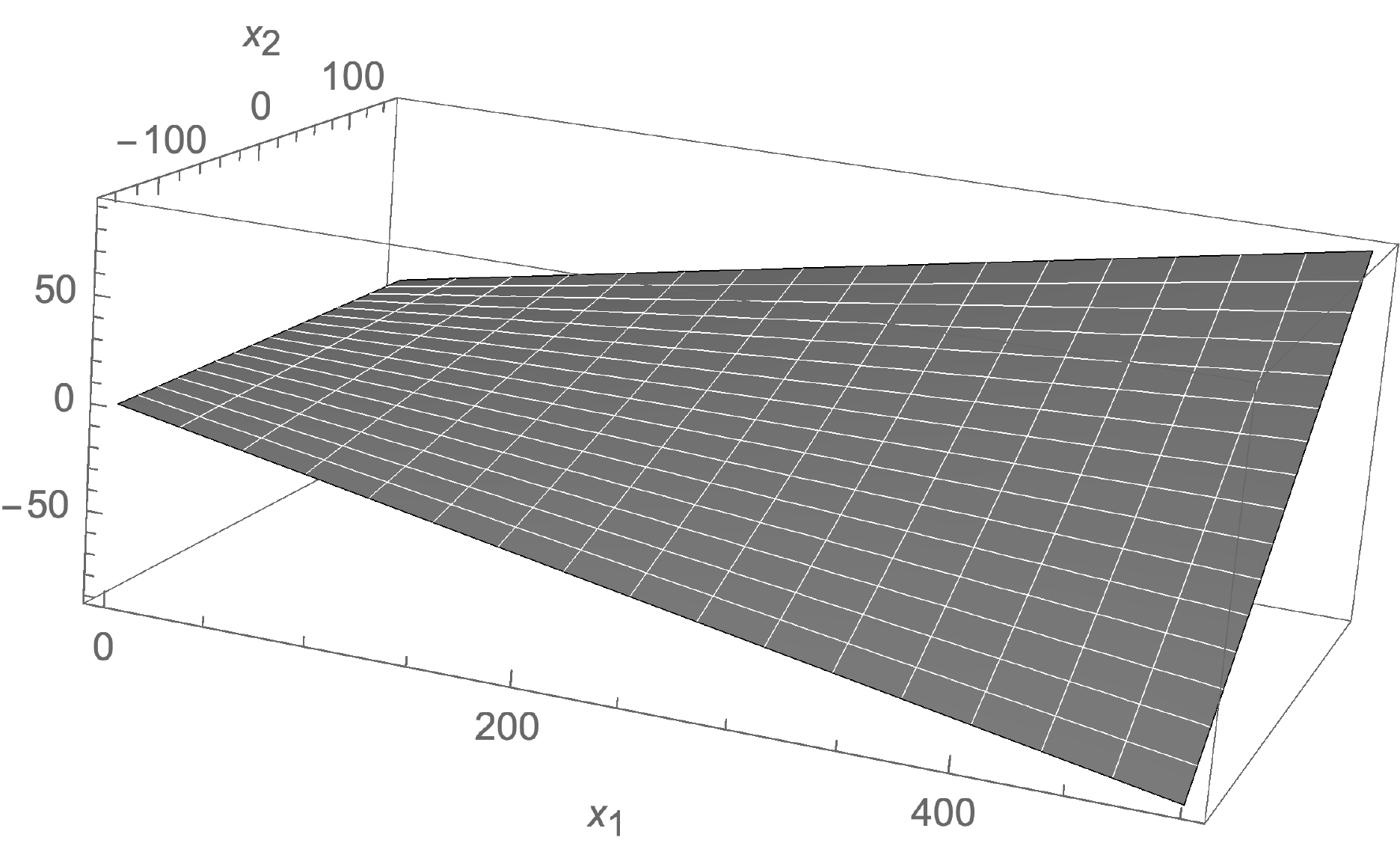}
\caption{The same coupled laminate  submitted to in-plane shear: the plate twists as an effect of coupling (units: mm).}
\label{fig:4}
\end{figure}

If now $N_1=N_2,N_6\neq0$, i.e. if the plate is submitted uniquely to in-plane shear, it will twist without bending. As an example, let us consider the same plate as above, now with dimensions 500 mm along $x_1$ and 300 mm along $x_2$, submitted to the in-plane shear $N_6=-2$ N/mm, which gives, eq. (\ref{eq:curvaturestrano}), $\kappa_6=2.4\times10^{-3}$ mm$^{-1}$. The deformation of the plate is shown in Fig. \ref{fig:4}.

 \subsubsection{Case $R_0^B=0.$} 

It is interesting  to analyze the case of the  condition $R_0^B=0$. In such a case we get
\be
\begin{split}
&t_0^A=t_0^D=\frac{T_0T_1-3{R_1^B}^2}{4T_0\left(T_0T_1-6{R_1^B}^2\right)},\\
& t_1^A=t_1^D=\frac{T_0}{16\left(T_0T_1-6{R_1^B}^2\right)},\\
&r_0^A=r_0^D=\frac{3{R_1^B}^2}{4T_0\left(T_0T_1-6{R_1^B}^2\right)},\\
&r_1^A=r_1^D=t_0^B=t_1^B=r_0^B=0,\\
&r_1^B=\frac{3{R_1^B}}{8\left(T_0T_1-6{R_1^B}^2\right)},\\
&\phi_0^A=\phi_0^D=\phi_1^B=\Phi_0^B.
\end{split}
\ee
So, $\Ac$ and $\Dc$ are no longer isotropic, but square symmetric, while $\Bc$ is $R_0$-orthotropic, like $\B$. Again fixing the frame in such a way that $\Phi_0^B=0$, it is
\be
\begin{split}
\Bc(\theta)=2\sqrt{2}r_1^B\left[
\begin{array}{ccc}
\sqrt{2}\cos2\theta&0&-\sin2\theta\\0&-\sqrt{2}\cos2\theta&-\sin2\theta\\-\sin2\theta&-\sin2\theta&0
\end{array}
\right],
\end{split}
\ee
Two cases of actions are interesting. The first one, is for $\theta=0$:
\be
\begin{split}
\Bc(\theta=0)=4r_1^B\left[
\begin{array}{ccc}
1&0&0\\0&-1&0\\0&0&0
\end{array}
\right],
\end{split}
\ee
so, if  $\N=(N_1,N_2,N_6)$, 
\be
\kappa_1=\frac{8}{h^2}r_1^B N_1,\ 
\kappa_2=-\frac{8}{h^2}r_1^B N_2,\
\kappa_6=0.
\ee
If $N_2=-N_1$,  $\kappa_1=\kappa_2$, but if  $N_1=N_2$,  then $\kappa_1=-\kappa_2$: like in the previous case, the Gaussian curvature is  negative, hence the deformed surface is made of hyperbolic points, and the plate bends as a minimal surface, because the mean curvature vanishes everywhere. Moreover, curvatures are insensitive to any in-plane shear. 

The second case is for $\theta=\frac{\pi}{4}$:
\be
\label{eq:Bstrange1}
\begin{split}
\Bc\left(\theta=\frac{\pi}{4}\right)=-2\sqrt{2}r_1^B\left[
\begin{array}{ccc}
0&0&1\\0&0&1\\1&1&0
\end{array}
\right],
\end{split}
\ee
so for $\N=(N_1,N_2,N_6)$ we have
\be
\kappa_1=\kappa_2=-\frac{4\sqrt{2}}{h^2}r_1^BN_6,\ \kappa_6=-\frac{4\sqrt{2}}{h^2}r_1^B(N_1+N_2).
\ee
Hence, for a pure in-plane shear, $N_6\neq0,N_1=N_2=0$, the plate will equally bend in the two directions $x_1$ and $x_2$, while for a pure stretch, $N_6=0$, the plate twists, unless $N_1=-N_2$, when no deformation is produced by coupling.

\subsection{Extension isotropy and bending orthotropy}
If only $\A$ is isotropic while $\D$ is orthotropic, like $\B$ and with the same axes and $\Phi_B=\Phi_D=0$, then $\Ac$ will remain isotropic if and only if
\be
\begin{split}
r_0^A\hspace{-1mm}=\hspace{-1mm}0\hspace{-1mm}\iff\hspace{-1mm}& 6 {R_1^B}^4 {T_0}+{T_1}\left({R_0^D}^2 {R_1^B}^2 -{R_0^B}^2 {R_1^D}^2\right)+\\
&+2 {R_0^B} {R_1^B} {R_1^D}{T_0} {T_1}\hspace{-1mm}-\hspace{-1mm}{R_1^B}^2 \left(3 {R_0^B}^2+{T_0}^2\right) {T_1}+\\
&+{R_0^D} \hspace{-1mm}\left({R_0^B}^2 {T_1}^2-6 {R_1^B}^4-2 {R_0^B} {R_1^B} {R_1^D} {T_1}\right)\hspace{-1mm}=\hspace{-1mm}0,\\
r_1^A\hspace{-1mm}=\hspace{-1mm}0\hspace{-1mm}\iff \hspace{-1mm}&R_1^B(T_1R_0^B-2R_1^BR_1^D)=0.
\end{split}
\ee
A set of sufficient conditions is  $\{R_1^B=0, {R_1^D}^2=T_1R_0^D\}.$ Though to get such a kind of laminates is not simple, but  possible through an optimization procedure, it is nevertheless interesting to consider them. After calculation, it is found that 
\be
\begin{split}
&t_0^A=\frac{T_0-R_0^D}{4\left(T_0^2-T_0R_0^D-3{R_0^B}^2\right)},\ t_1^A=\frac{1}{16T_1},\\
&t_0^D=\frac{T_0}{4\left(T_0^2-T_0R_0^D-3{R_0^B}^2\right)},\\
& t_1^D=\frac{T_0^2+T_0R_0^D-3{R_0^B}^2}{16T_1\left(T_0^2-T_0R_0^D-3{R_0^B}^2\right)},\\
&r_0^D=0,\ r_1^D=\frac{T_0}{8\left(T_0^2-T_0R_0^D-3{R_0^B}^2\right)}\sqrt{\frac{R_0^D}{T_1}}.
\end{split}
\ee
Hence, $\Dc$ is $R_0$-orthotropic. Moreover
\be
\begin{split}
\Bc=\left[
\begin{array}{ccc}
\Bc_{11}&\Bc_{12}&0\\-\Bc_{11}&-\Bc_{12}&0\\0&0&\Bc_{66},\\
\end{array}
\right],
\end{split}
\ee
with
\be
\label{eq:bstrano}
\begin{split}
&\Bc_{11}=\frac{3R_0^B(\sqrt{T_1R_0^D}-T_1)}{4T_1\left(T_0^2-T_0R_0^D-3{R_0^B}^2\right)},\\
&\Bc_{12}=\frac{3R_0^B(\sqrt{T_1R_0^D}+T_1)}{4T_1\left(T_0^2-T_0R_0^D-3{R_0^B}^2\right)},\\
&\Bc_{66}=\frac{3R_0^B}{2\left(T_0^2-T_0R_0^D-3{R_0^B}^2\right)}.
\end{split}
\ee
$\Bc$ is singular and $\Bc\neq\Bc^\top$. Then, also in this case  it cannot be represented by the  polar formalism for a tensor of $\mathbb{E}$la, eq. (\ref{eq:mohr4}) and  the converse of eq. (\ref{eq:bspecial}) must be used to compute the polar parameters, cf. \cite{vannucci10ijss}; we get
\be
\begin{split}
&t_0^B=t_1^B=t_3^B=r_2^B=0,\ \phi_0^B=\phi_1^B=0,\\
&r_0^B=\frac{3R_0^B}{4\left(T_0^2-T_0R_0^D-3{R_0^B}^2\right)},\\
&r_1^B=\frac{3R_0^B}{8\left(T_0^2-T_0R_0^D-3{R_0^B}^2\right)}\sqrt{\frac{R_0^D}{T_1}}.
\end{split}
\ee 
$\Bc$ is hence specially orthotropic, because $r_2^B=0$. However, its isotropic part is still null ($t_0^B=t_1^B=t_3^B=0$). 
It is interesting to notice that if only the condition $R_1^B=0$ is satisfied, then it will be $r_1^A=0, r_1^D\neq0,\phi_0^D=\phi_1^D=0$: $\Ac$ is square symmetric and $\Dc$ just ordinarily orthotropic. Concerning $\Bc$, it is still $\Bc\neq\Bc^\top$ and $r_2^B=0$, but now $t_0^B\neq0$: the isotropic part of $\Bc$ is no longer null, though the layers are identical. The situation for $\Bc$ does not change also if $R_1=0\Rightarrow R_1^A=R_1^B=R_1^D=0$, but now also $\Dc$ is square symmetric.

In all the cases, the matrix of $\Bc$ is like in eq. (\ref{eq:bstrano}); so if the plate is simply stretched, along any direction, it is $\kappa_1=-\kappa_2$: once more, the deformed plate is a minimal surface made of hyperbolic points.

\subsection{Square-symmetric coupled laminates}
Let us now consider a coupled laminate with $R_1^A=R_1^D=0$; we want to know when $r_1^A=r_1^D=0$. Proceeding in the same way, we get the unique condition
\be
R_0^BR_1^B\left[3{R_0^B}^2-(T_0-R_0^A)(T_0-R_0^D)\right]=0,
\ee
valid for $\A,\B$ and $\D$ coaxial, i.e. for $\delta_A=\delta_D=0$. It is interesting to remark that the square symmetry of $\A$ and $\D$ is preserved in compliance when also $\B$ is square symmetric but also when $\B$ is $R_0$-orthotropic. Let us consider these two cases separately.

\subsubsection{Case $R_1^B=0$.} 
If $R_1^B=0$ we get  
\be
\begin{split}
&{t_0^A= }\frac{{T_0} \left(T_0^2-{R_0^D}^2-3 {R_0^B}^2\right)}{4\mu},\\
&{t_1^A= }\frac{1}{16 {T_1}},\\
&{r_0^A= } {\frac{3 {R_0^B}^2 {R_0^D}+{R_0^A} \left({T_0}^2-{R_0^D}^2\right)}{4\mu}},\\
&{t_0^B= }\frac{3 {T_0}{R_0^B} ({R_0^A}+{R_0^D}) }{4\mu},\\
&{r_0^B= }{\frac{3{R_0^B} \left({R_0^A} {R_0^D}+{T_0}^2-3 {R_0^B}^2\right)}{4\mu}},\\
&{t_0^D= }\frac{ {T_0} \left({T_0}^2-{R_0^A}^2-3 {R_0^B}^2\right)}{4\mu},\\
&{t_1^D= }\frac{1}{16 {T_1}},\\
&{r_0^D= }{\frac{3 {R_0^B}^2{R_0^A} +{R_0^D}\left({T_0}^2-{R_0^A}^2\right)}{4\mu}},\\
&t_1^B=r_1^A=r_1^B=r_1^D=0,\\
& \phi_0^A=\phi_0^B=\phi_0^D=0,
\end{split}
\ee
with
\be
\begin{split}
\mu&= \left[3 {R_0^B}^2-(T_0-{R_0^A}) ({T_0}-{R_0^D})\right]\times\\
&\times\left[3 {R_0^B}^2-(T_0+{R_0^A}) (T_0+{R_0^D})\right].
\end{split}
\ee
The matrices of $\Ac,\Bc,\Dc$ are hence square symmetric and can be easily composed through eq. (\ref{eq:mohr4}). $\Bc$ is singular and symmetric and also in this case its isotropic part is not null:
\be
\begin{split}
\Bc(\theta)=\left[
\begin{array}{ccc}
\Bc_{11}(\theta)&\Bc_{12}(\theta)&\Bc_{16}(\theta)\\\Bc_{12}(\theta)&\Bc_{22}(\theta)&\Bc_{26}(\theta)\\\Bc_{16}(\theta)&\Bc_{26}(\theta)&\Bc_{66}(\theta)
\end{array}
\right].
\end{split}
\ee
with
\be
\begin{split}
&\Bc_{11}(\theta)=\Bc_{22}(\theta)=-\Bc_{12}(\theta)=t_0^B+r_0^B\cos4\theta,\\
&\Bc_{66}(\theta)=2(t_0^B-r_0^B\cos4\theta),\\
&\Bc_{16}(\theta)=-\Bc_{26}(\theta)=-\sqrt{2}r_0^B\sin2\theta.
\end{split}
\ee
So, for $\theta=0$, once more in the case of $\N=(N_1,N_2,N_6)$, we get
\be
\begin{split}
& \kappa_1=-\kappa_2=\frac{2}{h^2}(t_0^B+r_0^B)(N_1-N_2),\\
& \kappa_6=\frac{4}{h^2}(t_0^B-r_0^B)N_6.
\end{split}
\ee
Again, for a pure stretch, the plate will take the form of a minimal surface and if $N_1=N_2$, no curvature due to coupling will appear. Because the isotropic part of $\Bc$ is not null ($t_0^B\neq0$), it is not possible in this case to have a $\Bc$ with a shape like in eq. (\ref{eq:Bstrange1}). So, it cannot happen that the plate will simply twist, it will always bend, in all the cases: for $\theta=\frac{\pi}{8}$ it is, e.g.,
\be
\label{eq:Bstrange2}
\begin{split}
\Bc\left(\theta=\frac{\pi}{8}\right)=\left[
\begin{array}{ccc}
t_0^B&-t_0^B&-\sqrt{2}r_0^B\\-t_0^B&t_0^B&\sqrt{2}r_0^B\\-\sqrt{2}r_0^B&\sqrt{2}r_0^B&2t_0^B
\end{array}
\right]
\end{split}
\ee
and hence
\be
\begin{split}
&\kappa_1=-\kappa_2=\frac{2}{h^2}\left[t_0^B(N_1-N_2)-\sqrt{2}r_0^BN_6\right],\\
&\kappa_6=\frac{2\sqrt{2}}{h^2}\left[r_0^B(N_2-N_1)+\sqrt{2}t_0^BN_6\right].
\end{split}
\ee
We recall again  that a simple way to get $R_1^B=0$ when also $R_1^A=R_1^D=0$ is to use a square-symmetric basic layer, i.e. a layer with $R_1=0$. 

\subsubsection{Case $R_0^B=0$.}
If $R_0^B=0$ we get that $\Bc$ is singular and $\Bc\neq\Bc^\top$:
\be
\begin{split}
&{t_0^A= }\frac{T_0T_1-3{R_1^B}^2}{4(T_0-R_0^A)\left[T_1(T_0+R_0^A)-6{R_1^B}^2\right]},\\
&{t_1^A= }\frac{T_0+R_0^D}{16\left[T_1(T_0+R_0^D)-6{R_1^B}^2\right]},\\
&{r_0^A= } \frac{T_1R_0^A-3{R_1^B}^2}{4(T_0-R_0^A)\left[T_1(T_0+R_0^A)-6{R_1^B}^2\right]},\\
&{r_1^B= }{\frac{3{R_1^B} }{8\left[T_1(T_0+R_0^A)-6{R_1^B}^2\right]}},\\
&{r_2^B= }{\frac{3{R_1^B} }{8\left[T_1(T_0+R_0^D)-6{R_1^B}^2\right]}},\\
&{t_0^D= }\frac{ {T_0T_1-3{R_1^B}^2} }{4(T_0-R_0^D)\left[T_1(T_0+R_0^D)-6{R_1^B}^2\right]},\\
&{t_1^D= }\frac{T_0+R_0^A}{16\left[T_1(T_0+R_0^A)-6{R_1^B}^2\right]},\\
&{r_0^D= } \frac{T_1R_0^D-3{R_1^B}^2}{4(T_0-R_0^D)\left[T_1(T_0+R_0^D)-6{R_1^B}^2\right]},\\
&t_0^B=t_1^B=t_3^B=r_1^A=r_0^B=r_1^D=0,\\
& \phi_0^A=\phi_1^B=\phi_2^B=\phi_0^D=0.
\end{split}
\ee
So, $\Ac$ and $\Dc$ are square symmetric ($r_0^A=r_0^D=0$), while $\Bc$ is $R_0$-orthotropic ($r_0^B=0$) and has the shape
\be
\begin{split}
\Bc(\theta)=\left[
\begin{array}{ccc}
\Bc_{11}(\theta)&\Bc_{12}(\theta)&\Bc_{16}(\theta)\\\Bc_{21}(\theta)&\Bc_{22}(\theta)&\Bc_{26}(\theta)\\\Bc_{61}(\theta)&\Bc_{62}(\theta)&0
\end{array}
\right],
\end{split}
\ee
with
\be
\begin{split}
&\Bc_{11}(\theta)=-\Bc_{22}(\theta)=2(r_1^B+r_2^B)\cos2\theta,\\
&\Bc_{12}(\theta)=-\Bc_{21}(\theta)=2(r_1^B-r_2^B)\cos2\theta,\\
&\Bc_{16}(\theta)=\Bc_{26}(\theta)=-2\sqrt{2}r_2^B\sin2\theta,\\
&\Bc_{61}(\theta)=\Bc_{61}(\theta)=-2\sqrt{2}r_1^B\sin2\theta.
\end{split}
\ee
Hence, for $\theta=0$,
\be
\begin{split}
\Bc(\theta=0)=\left[
\begin{array}{ccc}
2(r_1^B+r_2^B)&2(r_1^B-r_2^B)&0\\-2(r_1^B-r_2^B)&-2(r_1^B+r_2^B)&0\\0&0&0
\end{array}
\right]
\end{split}
\ee
and for a given $\N=(N_1,N_2,N_6)$ it is
\be
\kappa_1=-\kappa_2=\frac{4}{h^2}\left[r_1^B(N_1+N_2)+r_2^B(N_1-N_2)\right].
\ee
Also in this case, for any set of in-plane actions the plate takes the form of a minimal surface.

For $\theta=\frac{\pi}{4}$,
\be
\begin{split}
\Bc\left(\theta=\frac{\pi}{4}\right)=-2\sqrt{2}\left[
\begin{array}{ccc}
0&0&r_2^B\\0&0&r_2^B\\r_1^B&r_1^B&0
\end{array}
\right]
\end{split}
\ee
and
\be
\kappa_1=\kappa_2=-\frac{4\sqrt{2}}{h^2}r_2^BN_6,\ \kappa_6=-\frac{4\sqrt{2}}{h^2}r_1^B(N_1+N_2),
\ee
i.e. any pure stretch ($N_6=0$) will twist the plate while a pure in-plane shear ($N_1=N_2=0$) will bend it.

\section{Conclusion}
The mathematical properties of the coupling tensors $\B$ and $\Bc$ have been considered in this paper, along with the mechanical consequences on the compliance tensors $\Ac$ and $\Dc$. It is apparent that the set of different situations is very large. What is interesting, especially for possible future applications, are the effects that coupling $\B$ has on $\Ac$ and $\Dc$, because these two tensors determine the response of a coupled laminate to a set of actions. This response can be very peculiar in some cases: on the one hand, the material symmetries of the stiffness tensors $\A$ and $\D$ can be obtained also for $\Ac$ and $\Dc$, but not automatically. On the other hand, some very special cases can be obtained too, e.g. $R_0$-orthotropy for $\Ac$ or $\Dc$, a type of orthotropy that can appear as rather theoretical but that becomes a real, possible situation produced by coupling on the compliance tensors. 

In this paper,  the  response of a coupled laminate to a mechanical action has been exclusively investigated; nevertheless, a topic of major interest is the response to a thermal action; in particular, it should be interesting to generalize the results of this paper to the case of thermally stable coupled laminates \cite{vannucci12joe1}, a set of laminates particularly important for applications. This is an ongoing research.


\subsection*{Acknowledgments} It is my pleasure to express my gratitude to Prof. B. Desmorat, University Paris-Saclay, for the fruitful discussions we had about the topic of this paper.

\bibliographystyle{plain}
\bibliography{Biblio.bib}

\end{document}